\documentclass[acmsmall]{acmart}

\AtBeginDocument{%
  }

\setcopyright{rightsretained}
\acmJournal{TECS}
\acmYear{2023} \acmVolume{1} \acmNumber{1} \acmArticle{1} \acmMonth{1} \acmPrice{}\acmDOI{}

\acmJournal{TECS}
\acmVolume{1}
\acmNumber{1}
\acmArticle{1}
\acmMonth{07}

\usepackage{graphicx}
\usepackage{xcolor}
\usepackage{multirow}
\usepackage[utf8]{inputenc}
\usepackage{algorithm} 
\usepackage{algpseudocode} 

\usepackage{color}
\usepackage{tabularray}

\definecolor{darkgreen}{RGB}{0,120,0}

\usepackage{stackengine}
\DeclareFontFamily{U}{mathx}{\hyphenchar\font45}
\DeclareFontShape{U}{mathx}{m}{n}{%
<-6> mathx5
<6-7> mathx6
<7-8> mathx7
<8-9> mathx8
<9-10> mathx9
<10-12> mathx10
<12-> mathx12
}{}

\DeclareSymbolFont{mathx}{U}{mathx}{m}{n}
\DeclareFontSubstitution{U}{mathx}{m}{n}

\DeclareMathSymbol{\bigovoid}{\mathop}{mathx}{"EC}

\usepackage{pifont}

\usepackage[super]{nth}
\usepackage{wrapfig}

\begin{document}


\title{MaGNAS: A Mapping-Aware Graph Neural Architecture Search Framework for Heterogeneous MPSoC Deployment}

\author{Mohanad~Odema}
\authornote{M. Odema and H. Bouzidi contributed equally to this research.}
\email{modema@uci.edu}
\affiliation{
  \institution{University of California Irvine}
  \city{Irvine}
  \country{USA}
}
\author{Halima~Bouzidi}
\authornotemark[1]
\email{Halima.Bouzidi@uphf.fr}
\affiliation{%
  \institution{Université Polytechnique Hauts-de-France}
  \city{Valenciennes}
  \country{France}
}
\author{Hamza~Ouarnoughi}
\email{Hamza.Ouarnoughi@uphf.fr}
\affiliation{
  \institution{Université Polytechnique Hauts-de-France}
  \city{Valenciennes}
  \country{France}
}
\author{Smail~Niar}
\email{Smail.Niar@uphf.fr}
\orcid{0000-0002-7550-484X}
\affiliation{%
  \institution{Université Polytechnique Hauts-de-France}
  \city{Valenciennes}
  \country{France}
}
\author{Mohammad~Abdullah~Al~Faruque}
\email{alfaruqu@uci.edu}
\orcid{0000-0002-5390-0497}
\affiliation{%
  \institution{University of California Irvine}
  \city{Irvine}
  \country{USA}
}

\thanks{This article appears as part of the ESWEEK-TECS special issue and was presented in the International Conference on Compilers, Architectures, and Synthesis for Embedded Systems (CASES), 2023.}


\begin{abstract}
Graph Neural Networks (GNNs) are becoming increasingly popular for vision-based applications due to their intrinsic capacity in modeling structural and contextual relations between various parts of an image frame.
On another front, the rising popularity of deep vision-based applications at the edge has been facilitated by the recent advancements in heterogeneous multi-processor Systems on Chips (MPSoCs) that enable inference under real-time, stringent execution requirements.
By extension, GNNs employed for vision-based applications must adhere to the same execution requirements. Yet contrary to typical deep neural networks, the irregular flow of graph learning operations poses a challenge to running GNNs on such heterogeneous MPSoC platforms. 
In this paper, we propose a novel unified \textit{design-mapping} approach for efficient processing of vision GNN workloads on heterogeneous MPSoC platforms. Particularly, we develop MaGNAS, a mapping-aware Graph Neural Architecture Search framework. MaGNAS proposes a GNN architectural design space coupled with prospective mapping options on a heterogeneous SoC to identify model architectures that maximize on-device resource efficiency.
To achieve this, MaGNAS employs a two-tier evolutionary search to identify optimal \textit{GNNs} and \textit{mapping} pairings that yield the best performance trade-offs. Through designing a supernet derived from the recent Vision GNN (ViG) architecture, we conducted experiments on four (04) state-of-the-art vision datasets using both (\emph{i}) a real hardware SoC platform (NVIDIA Xavier AGX) and (\emph{ii}) a performance/cost model simulator for DNN accelerators. Our experimental results demonstrate that MaGNAS is able to provide \textbf{1.57}$\times$ latency speedup and is \textbf{3.38}$\times$ more energy-efficient for several vision datasets executed on the Xavier MPSoC vs. 
the GPU-only deployment while sustaining an average \textbf{0.11\%} accuracy reduction from the baseline.

\end{abstract}

\ccsdesc[500]{Computing methodologies~Distributed computing methodologies}
\ccsdesc[500]{Computing methodologies~Neural networks}
\ccsdesc[300]{Computer systems organization~Embedded and cyber-physical systems}

\keywords{Graph Neural Networks, MPSoCs, HW-SW codesign, Edge Computing}


\maketitle

\section{Introduction}

Due to their inherent capacity in learning meaningful feature representations from non-Euclidean graph-structured data, the employment of Graph Neural Networks (GNNs) has extended beyond typical graph learning applications, e.g., molecular inference and social networks \cite{wu2020comprehensive}, to encompass the field of computer vision. 
By transforming an image structured as a regular grid of pixels into a graph, irregular and complex objects can be better captured by the more flexible graph-level features generated throughout the model architecture.
As such, recent works employing GNNs to operate on this generalized form of image data have demonstrated remarkable successes across a variety of visual tasks, e.g., object detection and image classification \cite{han2022vision, wang2019dynamic, yan2018spatial, yang2018graph}. In fact, the application of GNNs has been further studied for more nuanced visual-based tasks in critical application settings, such as collision prediction in self-driving vehicles \cite{yu2021scene, malawade2022spatiotemporal}.

On a separate note, recent advances have seen a proliferation in multi-processor System-on-Chips (MPSoCs) architectures that can balance the low-latency and energy efficiency requirements of compute-intensive workloads. For instance, commercial SoC platforms, such as the Nvidia Xavier \cite{agx} and Tesla FSD \cite{talpes2020compute}, have successfully integrated a variety of proven hardware computing units (CUs) and industrial IPs on a single chip to achieve said purpose. Other platforms, such as Xilinx Versal \cite{gaide2019xilinx}, enable even more flexibility in SoC solution development by supporting customized hardware design choices. Through such advanced platforms, deep learning-based vision modules can be run effectively in an edge computing setting to meet stringent application requirements such as object detection for autonomous driving \cite{lin2018architectural}. 
By extension, any consideration for applying GNNs in these vision modules under the embedded deployment setting must ensure that the execution constraints are still satisfied. However, this objective is challenging, considering the discrepancy between the GNN workloads and the underlying hardware in the SoC. That is, contrary to the dense, regular workloads of typical DNNs, GNNs are characterized by an \textit{irregular, multiphase sparse-dense} computational flow \cite{garg2022understanding}. Particularly, this irregularity emanates from the repeated sequence of \textit{Aggregation} and \textit{Combination} phases. The former employs a message-passing algorithm for feature exchange between graph vertices, exhibiting sparse kernels with random memory access patterns. The latter constitutes typical multi-layer perceptron (MLP) layer(s) for feature transformation, exhibiting dense kernels and regular access patterns. As such, the complication arises as neither the architecture of typical CUs (e.g., GPU) nor that of conventional accelerators (e.g., DLA) is designed to efficiently support this unique execution sequence.

Naturally, considerable research works have dedicated efforts to design customized GNN accelerator architectures that can support the \textit{multi-phased} computational flow \cite{chen2021dygnn, you2022gcod, stevens2021gnnerator, yan2020hygcn, auten2020hardware, kiningham2022grip}. Generally, the approach entailed a \textit{hybrid} architecture comprising specialized computing engines to accelerate each of the two phases separately. Unfortunately, these designs are not flexible enough to be consolidated into standard MPSoCs. On the one hand, this is attributed to the fact that GNNs belong to a relatively nascent, rapidly-evolving field in which customized accelerator architectures may not support running newer generations of graph learning operations and models. On the other hand, physical restrictions and low-power requirements of critical embedded computing platforms at the edge \textit{restrict} the integration of specialized hardware CUs onto the SoC to the components that best serve the desired target applications -- as in how DLAs are integrated in the AGX Xavier SoC as they support a broad class of applications which employ typical DNN workloads. 

As GNNs continue to become increasingly popular, the challenges of their deployment onto embedded platforms are due to be seen in a new light. In addition to implementing customized accelerator architectures, another research direction is to investigate what optimization opportunities exist -- on both the hardware and algorithmic levels -- to alleviate the deficiencies of GNNs' computational flow when deployed on conventional CUs. Researchers in \cite{garg2022understanding} have assumed this perspective by characterizing the design space of \textit{dataflow} choices for running GNNs on conventional re-configurable spatial accelerators, where they studied the costs and benefits of adopting various dataflows for GNNs. In that same spirit, we also believe there are ample optimization opportunities through characterizing the combined design space of \textit{SoC mapping options} and \textit{GNN architectural parameters} together. In the context of GNNs for vision applications, two considerations motivate this hypothesis: 
(\emph{i}) Heterogeneous MPSoCs naturally offer \textit{pipelining parallelism} opportunities, presenting options to run GNN kernels of diverse characteristics on different CUs to potentially yield better performance benefits. (\emph{ii}) the recently proposed VisionGNN (ViG) architecture \cite{han2022vision} offers to transform an image frame into a graph by dividing it into equally-sized patches and constructing a graph out of them to be processed by the model. As will be detailed later, the key advantage of this scheme is that it enables leveraging graph-level features while maintaining a consistent, dense structure for any graphed image throughout the GNN model, which is more amenable to CUs than sparse graphs of \textit{inconstant} dimensions.

\begin{figure}
\centering
    \includegraphics[width=\textwidth]{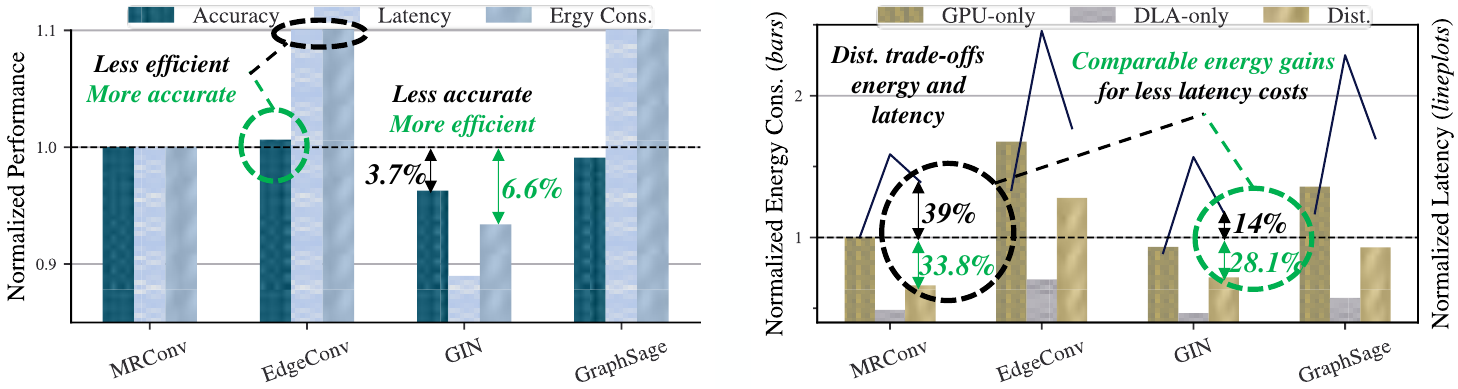}
    \caption{Comparing ViG model variants \cite{han2022vision} with different graph learning operators when trained on the Oxford-Flowers dataset and deployed onto the NVIDIA Jetson AGX Xavier SoC. All values are normalized by the baseline performance evaluations incurred by the original ViG with MRConv layers when fully deployed onto the GPU only. The \emph{left} figure shows how performance characteristics differ from one variant to the other regarding accuracy, latency, and energy consumption. The \emph{right} figure illustrates how distributed mapping strategies across the GPU and DLA can yield different latency-energy trade-offs.}
    \label{fig:motiv}
\end{figure}

\subsection{Motivational Example}

In Figure \ref{fig:motiv}, we showcase the potential performance trade-offs as offered by the \textit{architectural} and \textit{mapping} optimization spaces for a vision GNN model when deployed onto a heterogeneous SoC. In this example, the backbone GNN architecture is the ViG-S \cite{han2022vision}, the target platform is the NVIDIA Xavier AGX SoC, and the models are trained on the Oxford-Flowers image dataset. Given how the ViG belongs to the Graph Convolutional Network (GCN) class of GNNs, we construct three (03) additional variants of the baseline ViG with different GCN operators. Specifically, the original ViG architecture employs the \texttt{Max\text{-}Relative} Graph Conv (\texttt{MRConv}) graph operation throughout the entirety of its model, whereas the variants employ other GCN layer types, namely \texttt{EdgeConv}, \texttt{GIN}, and \texttt{GraphSage}. 
After training the ViG variants, we characterize their accuracy, latency, and energy consumption scores relative to the original \texttt{MRConv} ViG variant when deployed onto the NVIDIA platform. In the \textit{left} Figure, we can observe some performance trade-offs from varying this \textit{singular} GNN architectural setting, i.e., the GCN layer operator. For instance, the \texttt{EdgeConv} ViG variant can achieve slightly higher accuracy (0.69\% more) than the \texttt{MRConv} one at the expense of a considerable increase in latency and energy consumption. Contrarily, the \texttt{GIN} operation is 6.6\% more energy-efficient than \texttt{MRConv} at the expense of a 3.7\% decrease in accuracy. Though there is no clear dominance for one variant over the other, this analysis sheds light on the potential performance trade-off gains from optimizing the architectural design parameters. These gains can be further compounded when considered alongside feasible deployment options. In these first experiments, only the GPU component of the SoC was used as the target deployment hardware.

In the \emph{right} Figure, we showcase how additional performance trade-offs are attained considering the various deployment options for the ViG variants on the SoC. In this example, the considered options are \textit{standalone} deployment on either the GPU or DLA components or \textit{distributed} deployment across the two. We remark that the distributed deployment options follow the mapping strategies for GNN processing workloads provided by our optimization engine, detailed in a later Section. From the Figure, the straightforward observation is that for every ViG architecture, \textit{standalone} GPU deployment is the option with the fastest execution speeds, \textit{standalone} DLA deployment is the most energy-efficient alternative, and the distributed option compromises between the two. However, a more interesting perspective on mapping optimizations can be taken when considered part of a broader design problem. That is, combining both the \textit{architectural} and \textit{mapping} optimizations to achieve better performance trade-offs compared to performing optimizations for each design space in isolation. For instance, assume a designer's primary objective is to improve the ViG's energy efficiency while incurring minimal execution slowdown. From a pure resource efficiency perspective, a \textit{distributed} mapping strategy for the \texttt{GIN} \textit{architectural} variant can be more beneficial than directly distributing the original \texttt{MRConv} ViG workloads since the former achieves comparable energy efficiency gains to those of the latter (28.1\% to 33.8\%) at the expense of reduced latency costs (14\% to 39\%). Still, the caveat remains that the \texttt{GIN} variant is less accurate than the original ViG, and the question becomes \textit{how can we better characterize this combined architecture-mapping design space to attain better performance trade-offs for vision GNNs given the target task and SoC platform.}

\subsection{Novel Contributions}
In light of the above challenges, we list the key novel contributions of this paper:
\begin{itemize}
    \item  We study how vision GNNs can leverage distributed deployment across multiple CUs for performance efficiency when deployed onto a heterogeneous SoC.
    \item We present \textbf{MaGNAS}, a \underline{M}apping-\underline{a}ware \underline{G}raph \underline{N}eural \underline{A}rchitecture \underline{S}earch Framework for \textit{co-optimizing} the design of vision GNN (ViG) architectures and their SoC mappings.
    \item {MaGNAS} first contributes a self-contained framework for designing ViG supernets to characterize their search space of GNN-based architectural design choices.
    \item To specify the mapping problem, we derive a system model that characterizes the distributed deployment of GNNs onto heterogeneous SoCs and the incurred performance overheads.
    \item To identify optimal \textit{ViG architecture-mapping} pairs, MaGNAS solves a bilevel optimization problem via a two-tier evolutionary search algorithm of two optimization engines: an \textit{outer} engine to optimize GNN architectural design choices; an \textit{inner} engine to identify optimal mapping strategies for ViG workloads onto heterogeneous CUs.
    \item We conduct extensive experiments, in-depth analysis, and ablation studies on MaGNAS using a real MPSoC platform and hardware simulator on four (04) state-of-the-art vision datasets. Our findings have demonstrated the superiority of MaGNAS in designing and mapping ViG architectures onto heterogeneous CUs and its effective scaling capabilities on increasing levels of problem complexity. On the Nvidia Xavier SoC, MaGNAS provided on average \textbf{1.57}$\times$ latency speedup and \textbf{3.38}$\times$ more energy gains than the GPU-only deployment while sustaining an average \textbf{0.11\%} accuracy drop from the baseline.
  
\end{itemize}

\section{A Primer on Vision Graph Neural Network (ViG)}\label{sec:primer}

We briefly describe the main constituents of the ViG architecture \cite{han2022vision}, which pioneered a generic approach for graph-based image processing through modeling raw input images as graph structures.

\textbf{Graphing Image Data Structures.}
The ViG operates on images modeled as graphs of patches. A $W \times H \times C$ image is first partitioned into $N$ patches of dimensions $W'\times H' \times C'$. Each patch's dimensions can be viewed as a single feature vector $x_i \in \mathbb{R}^{D}$ where $D = W' \times H' \times C'$. To construct the graph, a node $v_i$ is assigned to each patch, forming an unordered set of $N$ nodes $\mathcal{V}=\{v_1, v_2,\dots, v_N\}$ associated with the corresponding set of feature vectors $X = \{x_1, x_2,\dots, x_N\}$, where $x_i$ can be called the \emph{feature embedding} of vertex $v_i$. To build graph edges, $K$ edges are constructed for each $v_i$ based on the $K$ nearest vertices in its neighborhood $\mathcal{N}(\mathcal{V})$, that is, for every $v_j \in \mathcal{N}(\mathcal{V})$, an edge $e_{ji}$ is constructed from $v_j$ to $v_i$. Finally, the full graph structure of the image is given by $\mathcal{G}(\mathcal{V}, \mathcal{E})$, which can be inputted into the ViG model for processing.

\textbf{Graph Processing Layer.}
Describing a graph through its features, $\mathcal{G} = G(X)$ s.t. $X\in\mathbb{R}^{N\times D}$, a typical GCN layer operation on $\mathcal{G}$ can be represented by the following abstract formula:
\begin{equation}
\mathcal{G'} = Combine(Aggregate(\mathcal{G}, W_{agg}), W_{comb})
\end{equation}
where $\mathcal{G}$ is processed through an \textit{aggregation} and a \textit{combination} stages of the GCN layer. $W_{agg}$ and $W_{comb}$ resemble the respective learnable weights of each stage. The \textit{aggregation} stage employs a feature exchange procedure in which every node $v_i$ receives features $x_j \in \mathcal{N}(x_i) s.t. i\neq j$ from its neighboring nodes and aggregates them to provide $x_i'$. The \textit{combination} stage involves further treatment of features $x_i'$ (as through an MLP layer) to obtain refined representation $x_i''$. We remark that for each of the two stages, a variety of operations can be employed (e.g., aggregation through sum, max-relative, mean), which correspond to the variety of GCN layer types existing in the literature (e.g., \texttt{GraphSage}, \texttt{GIN}, etc.). Lastly, The resulting output feature set from both stages, $X'$, is used to construct the output graph $\mathcal{G'} = G(X')$.

\textbf{Grapher and FFN Modules.} To enrich feature representation, graph processing layers can be interleaved with typical DNN layers in a GNN model. As such, the standard ViG architecture comprises a stack of two basic building blocks: \textit{Grapher} and \textit{Feed Forward Network (FFN)} given by: 
\begin{equation}
    L^{Grapher} = l^{post} \circ l^{comb} \circ l^{agg} \circ l^{pre}, \;\;\;\;\;\; L^{FFN} = l^{fc_2} \circ l^{fc_1} \label{eqn:layerwise}
\end{equation}
The \textit{Grapher} comprises at its core the GCN layer with its aggregation, $l^{agg}$, and combination, $l^{comb}$, operations, injected between two linear layers, namely \textit{pre-processing}, ($l^{pre}$), and \textit{post-processing}, $l^{post}$, layers, to promote feature diversity. The \textit{FFN} block constitutes two fully connected layers that further elevate feature capacity, $l^{fc_1}$ and $l^{fc_2}$.
For every GCN or fully-connected layer in either module, non-linear activation and batch normalization operations are applied. From here, every \textit{Grapher} can be followed by an optional \textit{FFN} to form the ViG block, and the sequence of ViG blocks form the ViG backbone architecture.

\section{System Model and Problem Formulation}
\label{sec:sys_model}

In this section, we model the mapping problem of GNN kernels onto heterogeneous SoC CUs. Then, we derive a formulation for the global design-mapping bi-optimization objective. 

\subsection{System Model for mapping GNNs onto Heterogeneous SoCs}

\subsubsection{GNN Workload Characterization}
Let a standard GNN model architecture, $\alpha$, be formally described as a sequence of $n$ computing blocks as follows:
\begin{equation}
\alpha = L_{n} \circ L_{n-1} \circ \cdots \circ L_{1}, \; s.t. \; L_i \neq L_{i-1},\; L_i \in \{L^{FFN}, L^{Grapher}\}, \; L^{FFN} \in \{L^{FFN}, \phi\} \; \forall 1 \leq i \leq n \label{eqn:GNN}
\end{equation}
where each GNN computing block $L_{i}$ can either be the \textit{Grapher} or \textit{FFN} blocks as defined in the previous section, denoted by $L^{Grapher}$ and $L^{FFN}$, respectively. The condition ensures that each $L^{Grapher}$ block can be succeeded by an optional $L^{FFN}$ block.

Let $X_j$ be the input graph-level features for block $L_j \in \alpha$. Then, the output feature embedding vector, $X_{j+1}$, can be obtained as:
\begin{equation}
    X_{j+1} = L_j(X_j) \;\; s.t. \;\; x_k^j \in \mathbb{R}^{D'} \; \forall \; x_k^j \in X_j \label{eqn:graph}
\end{equation}
where the condition ensures that feature embedding dimensions remain consistent throughout each computing block within the GNN. That is the feature embedding for $x_k^j$ (the $k^{th}$ node within the graph representation at the $j^{th}$ block) retains the same $D'$ dimensions before and after being processed through block $L_j$. This consistency in the feature embedding dimensions is typical of GNNs as it preserves the integrity of graph operations with regards to feature aggregation from farther nodes across multiple consecutive layers and facilitates supporting residual and dense connections
\cite{you2020design}. Note that $D'$ can either be equivalent to $D$ or a downsampled version of it as some architectures (e.g., Pyramid in \cite{han2022vision}) can include additional downsampling layers in-between stacks of computing blocks to promote abstract feature learning.

Let $\mathbb{CU} = \{\mathcal{CU}_1, \mathcal{CU}_2, \cdots, \mathcal{CU}_M\}$ be the set of available computing units within a heterogeneous MPSoC with varying degrees of support for DNN and graph operations. 
Considering a \textit{blockwise} granularity, we can define a mapping vector, $m$, to characterize the workload distribution for each GNN computational block as follows:
\begin{equation}
m = [\pi_1, \pi_2, \cdots, \pi_n], \;\; s.t. \;\; \pi_i \in \mathbb{CU} \; \forall \; 1 \leq i \leq n \; \vert \; support(\pi_i, L_i) == True \label{eqn:M}
\end{equation}
where each entry $\pi_i$ in $\mathbb{M}$ describes the mapping assignment of $L_i$ onto a computing unit $\mathcal{CU}_m \in \mathbb{CU}$ as long as this corresponding $\mathcal{CU}_m$ hardware supports running $L_i$.
 
\subsubsection{Performance Modelling.} For a mapping strategy $m$, the total latency and energy consumption overheads, $T_{total}$ and $E_{total}$, experienced by a GNN model when deployed in a distributed, pipelined fashion can be modeled as the sum of the overheads incurred by its individual blocks:
\begin{align}
    T_{total}(m) = \sum_{i=1}^{n} T_i(m), \;\; s.t. \;\; T_i(m) = \tau_i^{comp} + \mathbb{I}[\pi_{i-1}\neq\pi_i]\cdot\tau_i^{in} + \mathbb{I}[\pi_{i}\neq\pi_{i+1}]\cdot\tau_i^{out} \\ 
    E_{total}(m) = \sum_{i=1}^{n} E_i(m), \;\; s.t. \;\; E_i(m) = e_i^{comp} + \mathbb{I}[\pi_{i-1}\neq\pi_i]\cdot e_i^{in} + \mathbb{I}[\pi_{i}\neq\pi_{i+1}]\cdot e_i^{out}
\end{align}
where the $\tau_i^{comp}$ and $e_i^{comp}$ are the respective computational latency and energy consumption experienced by $L_{i}$ given its corresponding mapping, $\pi_i$. $\tau_i^{in}$ and $\tau_i^{out}$ are the latency overhead sustained when loading and writing back graph features from and to the \textit{shared system memory} on the SoC, respectively. The indicator function $\mathbb{I}[\cdot]$ evaluates to 1 only when the associated condition is met; that is, no transmission overhead penalties are sustained between two consecutive layers when they are both assigned the same computing unit. For the energy formula, the same logic of notation applies for every layer $L_i$.

\subsubsection{Mapping Problem Formulation} Define $P(m) = f(T_{total}(m), E_{total}(m))$ to be a combined evaluation function for a mapping configuration $m$. Let $\mathbb{M}$ be the set of feasible mapping configurations. Then, we can formulate the mapping objective function for an architecture $\alpha$ deployed on a heterogeneous SoC platform as follows:
\begin{equation}
   m^* = \max_{m \in \mathbb{M}} P(m), \;\; s.t. \;\; T_{total} < T_{TRG},\; E_{total} < E_{TRG} \label{eqn:map_obj}
\end{equation}
where the goal is to identify an optimal mapping strategy, $m^*$, for $\alpha$ such that performance objective function $P$ is maximized with respect to latency and energy under user-specified constraints on latency and energy consumption, $T^{TRG}$ and $E^{TRG}$, respectively.

\subsection{Nested Search Formulation} As the application of graph learning on embedded hardware is a relatively nascent field, the lack of standardization in GNN architectures for edge deployment settings adds another dimension to this design optimization problem. Together with the mapping formulation derived above, a natural question arises as follows: \textit{Given an awareness of the ideal mapping strategy for a GNN onto a heterogeneous MPSoC, can we leverage this information to guide further architectural design optimizations such that the target task accuracy and resource efficiency are enhanced?}

In light of this proposition, we refine our formulation to an \textit{architecture-mapping} co-optimization problem, where the goal is to identify the optimal set of design choices for the GNN architecture and its mapping strategy. Since a Cartesian product of their combined search parameters can result in an enormous search space, we designate two separate subspaces to be managed through a bi-level optimization approach as follows: a) GNN architecture subspace ($\mathbb{A}$); which describes the set of architectural design choices associated with the GNN model, and b) Mapping subspace ($\mathbb{M}$); specifying the possible distributed mapping options given the underlying CUs. Through this designation, mapping choices become conditioned on architectural choices, which promotes the generality of this approach. Formally, the nested optimization formulation can be given as follows:
\begin{align}
    \alpha^{*} = \max_{\alpha \in \mathbb{A}} \psi[Acc(\alpha), P(m^*\vert \alpha, \mathbb{CU})] \label{eqn:outer} \\ s.t. \; m^{*} = \max_{m\in\mathbb{M}}P(m\vert \alpha, \mathbb{CU}) \label{eqn:inner}
\end{align}
where the outer optimization equation targets identifying the optimal set of GNN architectural parameters, $\alpha^{*}$, that yield the best scores on a combined function, $\psi$, of both the accuracy, $Acc(\cdot)$, and performance efficiency $P(\cdot)$. Evaluation of $P(\cdot)$ is contingent upon the results from the inner optimization equation. That is, energy and latency performance evaluations used for scoring a candidate architecture, $\alpha$, are those obtained for an optimal mapping strategy, $m^{*}$. Due to the conflicting nature of the involved objectives, the problem can be solved as a multi-objective optimization providing a Pareto-optimal set of solutions. For instance for the outer optimization objective, an architecture $\alpha^{*}$ is said to be Pareto-optimal iff for every objective $u \in U$:
\begin{equation}
    u_{k}(\alpha^{*}) \geq u_{k}(\alpha) \forall k,\alpha\; \text{and} \; \exists j: u_{j}(\alpha^{*}) > u_{j}(\alpha) \forall (\alpha) \neq (\alpha^{*}) 
\end{equation}

\section{MaGNAS Framework}

To solve the above GNN \textit{architecture-mapping} co-optimization problem, we present \textbf{MaGNAS}, a mapping-aware Graph Neural Architecture Search framework for heterogeneous SoC deployment. \textbf{MaGNAS} employs two phases: (\emph{i}) the construction and training of a ViG supernet to attain a design space of diverse GNN architectural design choices; (\emph{ii}) the development of a two-tier evolutionary search framework to \textit{identify} optimal \textit{architecture-mapping} pairings.

\subsection{Supernet Construction and Training} \label{subsec:supernet}

\begin{figure}
\centering
    \includegraphics[width=\textwidth]{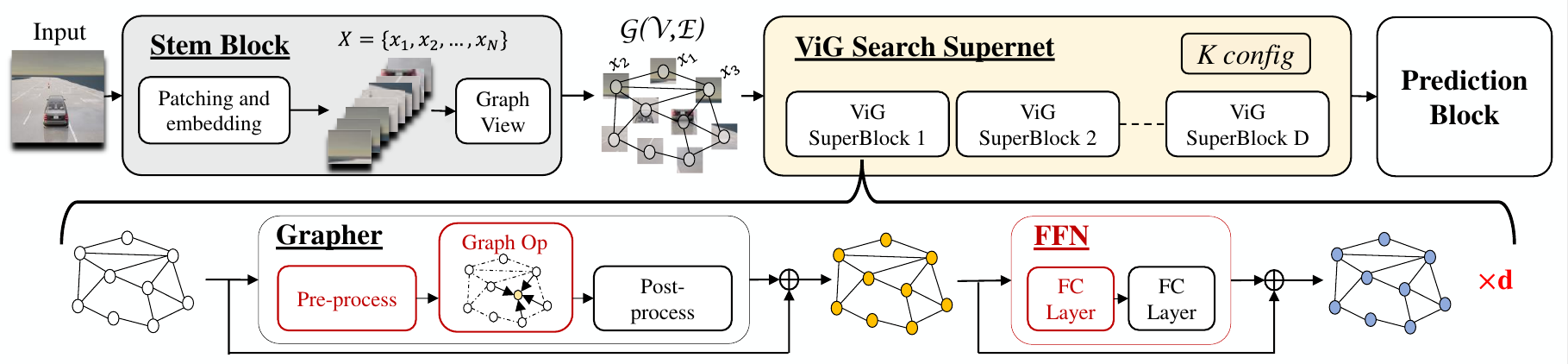}
    \caption{The ViG supernet implementation for MaGNAS co-search framework. The supernet comprises $D$ ViG search super blocks, each of which constitutes a sequence of $d_i$ Grapher and FFN computing modules. Architectural search parameters characterizing $\mathbb{A}$ subspace are highlighted in \textit{red} and detailed in the text.}
    \label{fig:vig_supernet}
\end{figure}

We extend the ViG architecture introduced in Section \ref{sec:primer} to construct a supernet of various design choices to characterize an architectural search space $\mathbb{A}$. Briefly, a \textit{supernet} represents a network of networks that can be trained simultaneously to facilitate providing diverse model designs for different deployment scenarios \cite{cai2019once}. In the context of ViGs, each subnet within a supernet is defined by a unique set of architectural parameter choices (e.g., choice of GNN layers, \#layers, etc.). Additionally, supernets entertain the property of \textit{weight-sharing}, meaning that during the supernet's training, weight updates for a candidate layer are applied and reused across all subnets that share that particular layer, which enables the simultaneous training of all subnets within it. Once the supernet is trained, a search algorithm can be employed to identify an ideal subnet that meets the target specifications.
The ViG supernet is illustrated in Figure \ref{fig:vig_supernet}, where the choice of architectural search parameters for $\mathbb{A}$ is based on observations from both related works \cite{you2020design, han2022vision, gao2020graph, you2022gcod} as well as from our initial experiments. The supernet construction is detailed in the following:

\subsubsection{ViG Superblocks} The backbone ViG-S architecture in \cite{han2022vision} comprises 16 computing blocks, each comprising a stack of a \textit{Grapher} and an \textit{FFN} module. On the one hand, characterizing $\mathbb{A}$ on a per-layer or a per-block basis can lead to an explosion in the search space, given the number and cardinality of various search parameters. Conversely, associating the parameters of $\mathbb{A}$ with the entire backbone restricts fine-grained architectural optimizations, not fully exploiting the power of diversified architectural settings at different model stages. As a compromise, we propose \textit{ViG superblocks} to characterize $\mathbb{A}$, where each $i^{th}$ superblock constitutes a collection of $d_i$ ViG blocks sharing the same design choices. Superblocks are inspired by the concept of neural computing blocks in popular architectures (e.g., ResNets), where the same architectural parameter value can be repeated for a stack of consecutive layers. Figure \ref{fig:vig_supernet} illustrates the composition of our ViG superblock and what architectural parameters are searchable within it.
The merits of the ViG superblocks are twofold: (\emph{i}) they balance the trade-off between architectural diversity and search space complexity; (\emph{ii}) They facilitate effective management of the depth parameter through $d_i$ while preserving key architectural features.

\subsubsection{$\mathbb{A}$ search parameters} \label{subsubsec:A} For each superblock $i$, we specify the following parameters to construct our architectural search space $\mathbb{A}$: 
\begin{itemize}
    \item \textit{The depth, }$d_i$, to indicate how many ViG blocks exist in the $i^{th}$ superblock $i$. 
    \item \textit{Grapher pre-processing} as a binary decision variable to indicate whether a pre-processing layer exists before every graph processing layer.
    \item \textit{Graph Op} to specify the graph operation employed throughout the $i^{th}$ superblock.
    \item \textit{FFN module} as a binary decision variable to indicate whether FFN modules should exist in this superblock.
    \item \textit{FC hidden layer dimension} to specify the size of the intermediate features in the FFN module.    
\end{itemize}
We do not include the Grapher's post-processing layer as part of $\mathbb{A}$ since, in the ViG backbone, it additionally contributes to maintaining the consistency of feature embedding dimensions.

\subsubsection{Supernet Training} We train the supernet for our target task using a combination of Cross-Entropy and knowledge distillation loss functions, where for the latter, we employ a pretrained model as a teacher for more representative training on soft labels' training \cite{yu2020bignas, bouzidi2023hadas}. This training is performed from scratch due to: (\emph{i}) The ViG is a relatively new GNN architectural concept, and the availability of pretrained weights is still limited, and (\emph{ii}) loading the exact pretrained model weights from the original ViG backbone \cite{han2022vision} can introduce a bias towards certain design choices during training. For instance, the original ViG architecture employed \texttt{MRConv} \textit{Graph Op} throughout the entirety of its graph processing layers. As such, loading their pretrained weights gives MRConv operations an edge over the remaining \textit{Graph Op} choices.

To train the supernet, we sample and train a set of subnets at each iteration. The choice of subnets is realized through 3 separate samplers following the Sandwich sampling rule \cite{yu2020bignas} as follows: 

\begin{itemize}
    \item \textit{Maximum Sampler:} sample the largest subnet from $\mathbb{A}$, that is, the one with the maximum depth and width (i.e., hidden dimension features).
    \item \textit{Minimum Sampler:} sample the smallest subnet from $\mathbb{A}$.
    \item \textit{Balanced Sampler:} sample a number of random subnets of different architectural features. 
\end{itemize}

This scheme enables improving the performance of all subnets within the search space simultaneously by pushing the upper and lower performance bounds with every iteration. Furthermore, given how numerous GNN architectures leverage a homogeneous structure, that is, one where the choice of the \textit{Graph OP} is kept consistent throughout the entire architecture, we modify the Maximum/Minimum samplers so that they sample architectures of maximal/minimal sizes, but constituting a randomly selected \textit{Graph Op} repeated throughout the model. This ensures training fairness by pushing the upper and lower boundaries of architectures of different graph operations and avoids inducing a bias towards specific implementations. 

\subsection{Nested Evolutionary Search: Outer Optimization Engine (OOE)}
\begin{figure}
\centering
    \includegraphics[width=\textwidth]{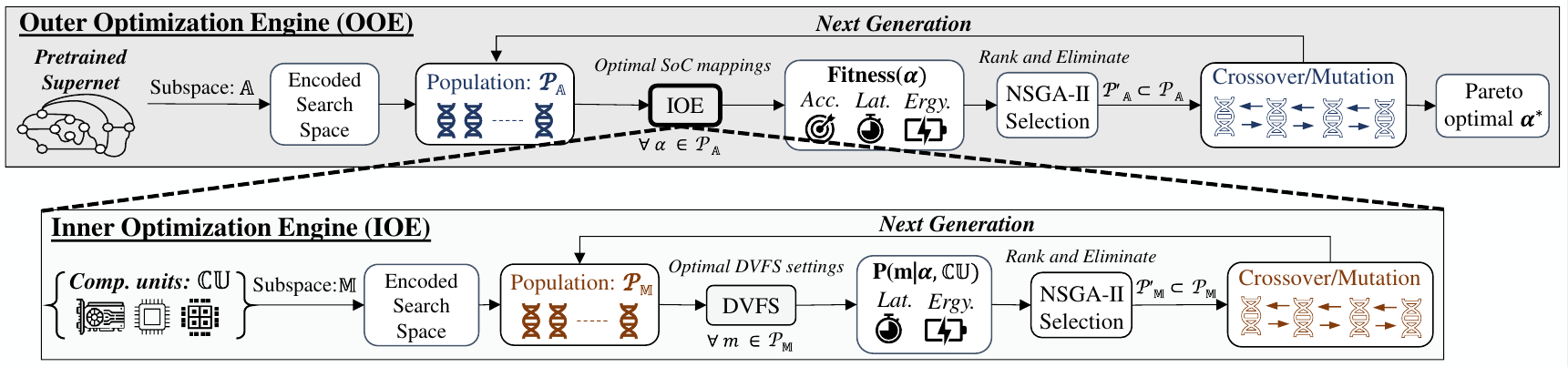}
    \caption{MaGNAS two-tier evolutionary search framework}
    \label{fig:evolution}
\end{figure}

In order to solve the bi-level \textit{architecture-mapping} optimization problem formulated in equations (\ref{eqn:outer}) and (\ref{eqn:inner}), we construct the two-tier evolutionary search framework illustrated in Figure \ref{fig:evolution} to identify optimal architecture-mapping pairings.
Briefly, an evolutionary search is a metaheuristic based on the concept of natural selection in biological evolution, where only the best individuals survive. Specifically, an evolutionary search works by creating a population of candidate solutions from a search space, evaluating each one, and propagating the top-performing solutions to the gene pool of subsequent generations. These solutions can then endure and undergo the genetic operations of mutation and crossover to contribute new derivative solutions for the following generations. This search paradigm is widely used in NP-hard problems to quickly retain optimal solutions while ensuring a broad exploration of gene diversity. In other words, an evolutionary search relies on updating a non-dominated solutions archive with every generation. Thus with each evolution, only new non-dominated solutions from the current population are added, and the newly-dominated ones in the archive are removed.

We first describe the Outer Optimization Engine (OOE), which employs a higher-level evolutionary algorithm whose purpose is to: (\emph{i}) search through the supernet to identify the most-promising GNN subnets and (\emph{ii}) rank candidate subnets according to their $Acc(\cdot)$ and $P(\cdot)$ evaluations. 

\subsubsection{Subspace $\mathbb{A}$ Description} 
By adopting a Once-For-All (OFA) NAS approach \cite{cai2019once}, the \textit{training} and \textit{search} stages within MaGNAS are decoupled, significantly reducing the search process overheads as once the supernet has been trained, its search subspace, $\mathbb{A}$, can be reused for the search to identify beneficial subnets. 
Accordingly, subspace $\mathbb{A}$ in the search stage is encoded as a sequence of 04 discrete vectors, each representing the architectural parameters for each ViG superblock listed in \ref{subsubsec:A}, facilitating the sampling of subnets as GNN architectural design candidates, $\alpha \in \mathbb{A}$.

\subsubsection{OOE Evolutionary Search}
The next step is to employ a search algorithm to solve the optimization objective in (\ref{eqn:outer}) by searching for optimal GNN architectural implementations, $\alpha^{*}$. Here, we implemented the NSGA-II evolutionary search algorithm to navigate through $\mathbb{A}$ and explore the subspace of viable design choices. Typically, the search algorithm is run for a pre-specified number of \textit{generations}, where a new population of candidate architectural designs, $\mathcal{P}^{g}_{\mathbb{A}}$, is sampled with every generation, $g$. Then, $\forall \alpha \in \mathcal{P}^{g}_{\mathbb{A}}$, a \textit{fitness} evaluation function, $F(\cdot)$, is applied as follows:
\begin{equation}
    F(\alpha) = f(Acc_{\alpha}, T_{\alpha}, E_{\alpha}) \label{eqn:OOE_rank}
\end{equation}
which scores every $\alpha$ based on its target task accuracy, latency, and energy consumption on the target platform denoted by $Acc_{\alpha}$, $T_{\alpha}$, and $E_{\alpha}$, respectively. $Acc_{\alpha}$ evaluation can be obtained directly by evaluating the $\alpha$ model predictive performance on the test dataset, whereas estimates of $T_{\alpha}$, and $E_{\alpha}$ are provided by the inner optimization engine based on evaluations of the ideal mapping strategy, $m^{*}$ (which will be detailed in the following subsection). Though we used for $F(\cdot)$ a weighted product function of the objective evaluations in our implementation, we kept its definition here abstract for generality. According to the fitness evaluation scores, every $\alpha \in \mathcal{P}^{g}_{\mathbb{A}}$ is ranked via the NSGA-II non-dominated sorting algorithm. Based on the rankings, an elimination process is initiated afterward to yield a population subset $\mathcal{P'}^{g}_{\mathbb{A}} \subset \mathcal{P}^{g}_{\mathbb{A}}$. Subset $\mathcal{P'}^{g}_{\mathbb{A}}$ then undergoes \textit{mutation} and \textit{crossover} operations to provide a new population $\mathcal{P}^{g+1}_{\mathbb{A}}$ for the following generation $g+1$. 
A uniform mutation is employed on the superblock level by sampling new depth, width, graph operators, etc., under a probability threshold of 0.4. The crossover is applied by randomly picking two individuals from the Pareto set and swapping their superblocks under a probability threshold of 0.5.
This iterative search continues until the search budget expires (e.g., a given total number of generations). At the last iteration, a Pareto-optimal set, $\{\alpha^{*} \vert m^{*}\}$, is provided. To provide some perspective based on our experiments, we sample 100 architectures for $\mathcal{P}^{g}_{\mathbb{A}}$ out of a total $\vert{\mathbb{A}}\vert \eqsim 2^{29}$ candidates. After fitness evaluations, we select a subset of 30\% from the top-ranked candidates as $\mathcal{P'}^{g}_{\mathbb{A}}$ for the following mutation and crossover processes.

\subsection{Nested Evolutionary Search: Inner Optimization Engine (IOE)}
To estimate $T_{\alpha}$ and $E_{\alpha}$ $\forall \alpha \in \mathcal{P}^{g}_{\mathbb{A}}$, we develop an Inner Optimization Engine (IOE) to specify an ideal mapping strategy of $\alpha$ onto the underlying SoC ($\alpha \rightarrow \mathbb{CU}$) and evaluate performance accordingly. 

\subsubsection{Subspace $\mathbb{M}$ Description} 
The mapping configuration, $m$, defined in equation (\ref{eqn:M}) reflects the encoded discrete vector within the IOE search space that characterizes potential mapping options for each \textit{Grapher} and \textit{FFN} modules from $\alpha$.
We also extend the specification of $m$ in the IOE to incorporate two further mapping options for the \textit{stem} and \textit{prediction} modules (see Figure \ref{fig:vig_supernet}). 

\subsubsection{IOE Evolutionary Search}
Given how the mapping decision space is at least $\vert \mathbb{CU} \vert ^ n$ (see equation (\ref{eqn:GNN})), a brute-force search to determine the ideal mapping, $m^{*}$, can be costly. As such, we implement another NSGA-II evolutionary algorithm in the inner optimization level to effectively explore mapping choices within $\mathbb{M}$ and identify the best candidates. Particularly, a population of mapping configurations, denoted by $\mathcal{P}_{\mathbb{M}}^{g}$, is sampled every generation $g$ by the search algorithm. Then for every $m\in\mathbb{M}$, a fitness evaluation function $P(\cdot)$ is applied as given in the below formula:
\begin{equation}
    P(m\vert \alpha, \mathbb{CU}) = (\frac{E_{\alpha}^{m}}{max \{E_{\alpha}^{\mathcal{CU}}\} })^{\gamma_1} \times (\frac{L_{\alpha}^{m}}{max \{L_{\alpha}^{\mathcal{CU}}\} })^{\gamma_2} \;\;\; \forall \mathcal{CU} \in \mathbb{CU} 
    \label{eqn:ioe_fitness}
\end{equation}
where $E_{\alpha}^{m}$ and $L_{\alpha}^{m}$ are the respective energy and latency sustained by $\alpha$ when its components are deployed onto the underlying hardware following a mapping strategy $m$. Each of these values is then normalized by the best \textit{standalone} deployment option from $\mathbb{CU}$, denoted here by $E_{\alpha}^{CU}$ and $L_{\alpha}^{CU}$, respectively. The reasons for this normalization are twofold: (\emph{i}) To ensure fairness when comparing various mapping options for $\alpha$; (\emph{ii}) To enforce achieving comparable, if not improved, performance scores over those obtained by the canonical standalone deployment options. For instance, if mapping the entirety of $\alpha$ onto a \textit{GPU} component is the best option with respect to latency, then all latency evaluations are normalized by $L_{\alpha}^{GPU}$. $\gamma_1$ and $\gamma_2$ are user-specified tunable hyperparameter values to enable prioritizing one performance objective or the other. 
For our experiments, we constructed accessible lookup tables by benchmarking computing blocks of varying architectural configurations onto the target CUs, allowing low-overhead estimations of latency and energy during the search.

Based on these evaluations, another non-dominated sorting algorithm is instantiated to rank mapping configurations, retaining the top-ranked configurations to provide population subset $\mathcal{P'}_{\mathbb{M}}^{g} \subset \mathcal{P}_{\mathbb{M}}^{g}$. Afterwards, subset $\mathcal{P'}_{\mathbb{M}}^{g}$ undergoes mutation and crossover to provide $\mathcal{P}_{\mathbb{M}}^{g+1}$ as the new population for the next generation. 
The mutation is uniformly applied by flipping the CU for each GNN computing block under a probability threshold of 0.4. The crossover is applied by randomly selecting two individuals from the Pareto set and interchanging their CUs mapping under a probability threshold of 0.8.
Once the search budget expires, $E_{\alpha}^{m^*}$ and $L_{\alpha}^{m^*}$ are returned as evaluations for the best configuration, $m^*$, to be used for $E_{\alpha}$ and $T_{\alpha}$ in the OOE, respectively. 

\subsubsection{Constrained Search} To support specifying $L_{TRG}$ and $E_{TRG}$ as search constraints during the search procedure as in equation (\ref{eqn:map_obj}), we designate an additional option for the selection procedure of the IOE non-dominated sorting algorithm to filter out mapping options from $\mathcal{P}_{\alpha}^{m}$ that do not conform to the pre-specified constraints, allowing only compliant mapping options to proceed to the next stage of mutation and crossover. If there were no compliant mappings, the standalone evaluations are returned for $E_{\alpha}$ and $T_{\alpha}$. In general, $L_{TRG}$ and $E_{TRG}$ can also be instated at the selection process of the OOE, where $\alpha$ architectures whose $E_{\alpha}$ and $T_{\alpha}$ do not meet target performance scores are eliminated from the population before the OOE's mutation and crossover stage.

\subsubsection{Performance Characterization} Generally, estimates of $E_{\alpha}^{m}$ and $L_{\alpha}^{m}$ for every $m \in \mathcal{P}_{\mathbb{M}}^{g}$ can be provided through a multitude of approaches (e.g., predictive models). As was shown in equation (\ref{eqn:graph}), the dimensional consistency of graph features offered throughout the ViG backbone has led to a tractable space of evaluation possibilities, enabling the construction of low-cost lookup tables to directly retrieve performance estimates of various architecture-mapping configurations. Simply put, the lookup tables are indexed by the architectural parameters of a computing block, $L_i$, and the $CU$ to whom it is mapped. By invoking the tables for every block in $\alpha$ given $m$, the performance overheads of each block can be aggregated to estimate the total $E_{\alpha}^{m}$ and $L_{\alpha}^{m}$. Although lookup tables work for our case, proxy prediction models can be more feasible for a different GNN architecture in which the graph features dimensions change as a result of inconsistent graph structures.

\subsubsection{DVFS Search Support} We also include the option to supplement $\mathbb{M}$ subspace with the configuration setting choices of dynamic voltage and frequency scaling (DVFS) features. Predominantly, numerous standard heterogeneous SoC components integrate this feature to support a diverse set of operational modes serving different execution contexts, as in to enable switching between \textit{low-power} and \textit{high performance} modes.
Here, to better capture the fine-grained effects of altering DVFS settings, we specify a DVFS search block in the IOE as a \textit{third} optional optimization level contingent upon the choices of $m$ and $\alpha$. This is convenient as the search space of the DVFS is small compared to $\mathbb{A}$ and $\mathbb{M}$ and does not incur as much search overhead. In typical real-time operational contexts, DVFS settings are kept the same across all the computing blocks of $\alpha$. This made a direct brute-force search through DVFS options sufficient to identify configurations that maximize the IOE fitness score in objective (\ref{eqn:ioe_fitness}). Formally, if we denote a single set of DVFS configuration settings as $\vartheta$ and the overall DVFS search space as $\Psi$, then the DVFS search objective is given as: 
\begin{equation}
    \vartheta^{*} = \max_{\vartheta \in \Psi} P(m \vert \alpha, \mathbb{CU}, \vartheta)
\end{equation}
where the performance evaluation of $m$ becomes also contingent upon the choice of $\vartheta \in \Psi$.

\section{Experiments}

In this Section, we conduct extensive experiments, in-depth analysis, and ablation studies using a real MPSoC platform and hardware simulation on four(04) state-of-the-art image classification datasets to assess the merit of MaGNAS in designing ViG architectures and mapping them onto heterogeneous CUs, as well as its ability to scale with an increasing degree of problem complexity.  

\subsection{Experimental Setup}

\subsubsection{Supernet Design.} \label{subsubsec:supernet}
We build our supernet on top of the ViG-S variant \cite{han2022vision} with 16 computing blocks, each a \textit{Grapher} and an \textit{FFN} block. We group every four (04) computing blocks into a \textit{ViG superblock}, and assign to each $K$ nearest neighbor values of 12, 16, 20, and 24, respectively, which enables aggregation of features from farther nodes with each superblock. 
To support dynamic width and depth configurations, we transform each ViG superblock into a \textit{slimmable} neural network following \cite{yu2018slimmable}.  
To support varying graph operations, we specify a dynamic graph processing layer in the \emph{Grapher} with four concurrent branches reflecting different GCN operational choices for \emph{Graph Op}: 1) \texttt{EdgeConv} \cite{wang2019dynamic}, 2) \texttt{GIN} \cite{xu2018powerful}, 3) \texttt{GraphSAGE} \cite{hamilton2017inductive}, and 4) \texttt{Max-Relative} GraphConv \cite{li2019deepgcns}. 
As mentioned in Section \ref{subsec:supernet}, the GNN search space also includes options to skip the \textit{Grapher}'s pre-processing layer and the entirety of the \textit{FFN} module throughout a given ViG superblock.

\begin{wraptable}{r}{0.6\textwidth}
  \centering
  \vspace{-5mm}
  \hspace{-4mm}
  \caption{Search space parameters for GNN architectures.}
  \vspace{-3mm}
  \label{tab:joint_search_space}
  \fontsize{9}{9}\selectfont
    \scalebox{0.69}
    {
        \begin{tblr}{
      row{2} = {c},
      row{8} = {c},
      row{11} = {c},
      cell{2}{1} = {c=3}{},
      cell{8}{1} = {c=3}{},
      cell{11}{1} = {c=3}{},
      hline{1-3,8-9,11-12,16} = {-}{},
    }
    \textbf{Decision variables}                                           & \textbf{Values}                                             & \textbf{Cardinality}  \\
    \textbf{Supernet Search Space ($\mathbb{A})$}                         &                                                             &                       \\
    Superblock depth (d)                                                 & \{2, 3, 4\}                                                 & 3                     \\
    Graph Op                                          & \{\texttt{Max\text{-}Relative}, \texttt{EdgeConv}, \texttt{GraphSAGE}, \texttt{GIN}\}                  & 4                     \\
    Skip pre-process (fc\_use)                                   & \{False, True\}                                             & 2                     \\
    Skip post-process (ffn\_use)                               & \{False, True\}                                             & 2                     \\
    FFN hidden features (w)                                               & \{96, 192, 320\}                                            & 3                     \\
    \textbf{Mapping Search Space ~\textbf{\textbf{\textbf{($\mathbb{M}$)}}} for NVIDIA Xavier AGX} &                                                             &                       \\
    Computing units                                         & \{GPU, DLA\}                                                & 2                     \\
    Mapping granularity                                                   & \{Stem, Grapher, FFN, Cls\}                          & $\mathcal{O}$(1.7$\times$10$^{12}$) \\
    \textbf{DVFS Settings Search space ($\Psi)$ for NVIDIA Xavier AGX}                          &                                                             &                       \\
    CPU clock frequency                                & \{1728MHz, 2265MHz\}                                        & 2                     \\
    GPU clock frequency                                & \{520MHz, 900MHz, 1377MHz\} & 3  
    \\
    EMC clock frequency                               & \{1065MHz, 2133MHz\} & 2  
    \\
    DLA clock frequency                                & \{1050MHz, 1395MHz\} & 2  
    \end{tblr}
        }
      \vspace{-4mm}
\end{wraptable}

\subsubsection{Datasets and Training.} 
We employ four (04) image classification datasets of CIFAR-10, CIFAR-100, Tiny-Imagenet, and Oxford-Flowers. To transform the images to graphs, images are first scaled to 224$\times$224$\times$3 resolution, and transformed through the \textit{Stem} block into a graph of nodes $N=196$, each of dimension $D=14\times14\times320$. The supernet training for each dataset is run for 150, 150, 250, and 250 
for each respective dataset in the order in which they were stated. The training is performed using an Adam optimizer with a momentum of 0.9, weight decay of 0.05, and dropout set to 0.2. We use cosine as a learning rate scheduler with an initial LR of 0.003 and batch size of 320 on a cluster of 20 GPUs of Nvidia RTX 2080 Ti (11 GB). 

\subsubsection{Evolutionary Search Settings.}
Table \ref{tab:joint_search_space} lists the search sub-spaces of $\mathbb{A}$, $\mathbb{M}$, and $\Psi$ designated within our optimization framework. 
For the optimization process, we fix the population size to 100 and 200 and the number of generations to 50, and 10 for the OOE and IOE, respectively. We adopt uniform mutation and crossover with respective probabilities of 0.8 and 0.4. We employ a dynamic encoding scheme 
in which the IOE evolutionary algorithm changes the size of the genome vector -for the mapping strategy encoding- according to the architectural parameters of the sampled GNN to avoid sampling meaningless decision variables (e.g., mapping choices for skipped FFN and FC-pre layers). Combining the OOE and IOE, we explored $\sim1.6\times10^6$ candidates of GNN architectures and deployment settings on an Nvidia Xavier AGX platform. The search process takes around $\sim$\textbf{1-2} GPU days to complete, depending on the complexity of the accuracy evaluation for each dataset.

\subsubsection{Hardware experimental settings}
\label{subsubsec:sim}
We evaluate our approach using two hardware experimental setups presenting a variety of computing units and architectural features: (i) NVIDIA Jetson AGX Xavier \cite{agx}, as a real target MPSoC platform; (ii) MAESTRO \cite{kwon2020maestro, kwon2019understanding}, as a hardware simulator tool. 

\noindent{\large \textcircled{\small 1}} \textbf{NVIDIA Jetson AGX Xavier: }
We employ the NVIDIA Jetson AGX Xavier MPSoC \cite{agx} as our primary experimental testbed. The platform is equipped with a high-performance Volta GPU of 512 GPU cores and 64 Tensor cores, and an energy-efficient DLA. We specify both components as the usable computing units of $\mathbb{CU}$ and characterize them as the feasible deployment options of $\mathbb{M}$. Both components share the same 16 GB 256 bits LPDDR4x 136,5 GB/s system memory and are orchestrated by the same CPU NVIDIA Carmel Arm 64 bits. To run workloads on GPU/DLA, we use the TensorRT 8.4 compiler running on top of CUDA 11.4 and cuDNN 8.3.2. As TensorRT is limited by the set of operations that can be executed on DLA, we consider this limitation in our performance characterization by enabling the \textit{GPU fallback} feature for the non-supported operations. 
The AGX Xavier also supports hardware reconfiguration of the clock frequencies of CPU, GPU, EMC, and DLA to emulate different hardware settings and power budgets, which we use to implement the DVFS search space $\Psi$. Unless otherwise stated, performance evaluations in our experiments are performed under the high-performance DVFS setting (MaxN).

\noindent{\large \textcircled{\small 2}} \textbf{MAESTRO: }
For the hardware scalability analysis, we leverage the MAESTRO tool \cite{kwon2020maestro, kwon2019understanding} to simulate a use-case of an SoC with three (03) heterogeneous CUs, where the heterogeneity is expressed by varying the dataflow configuration on each accelerator given how different neural network workloads exhibit different affinities towards dataflow choices for maximizing performance efficiency. For example, a weight stationary dataflow (like \textit{kcp\_ws} from MAESTRO and that of the DLA accelerator in the Nvidia Xavier) maximizes filter weights' reuse which is useful for layers whose same filters are used to compute multiple outputs, limiting the number of times weights need to be fetched from the main memory and improving energy efficiency in the interim \cite{chen2016eyeriss}. We use the native dataflows in MAESTRO of \textit{kcp\_ws}, \textit{ykp\_os}, and \textit{dpt} for our 3 CUs, which for simplicity, we denote by \textbf{DSA-k}, \textbf{DSA-y}, and \textbf{DSA-d}. We also use for this experiment the PyramidViG-M architecture detailed in the following. 

\subsubsection{Baselines.}\label{subsubsec:base}

The efficacy of our approach is assessed regarding the following GNN architectural and hardware mapping baseline:

\noindent{\large \textcircled{\small 1}} \textbf{GNN architectures baselines:} These include the original isotropic ViG-S model in \cite{han2022vision} as well as its variants by altering \textit{Graph Op} (i.e., the GCN operation) where the \textit{Graph Op} remains consistent across all the layers. Specifically, we identify the baselines by their recurring \textit{Graph Op} operation: 1) \textbf{b0}: ViG-S/Max-Relative, 2) \textbf{b1}: ViG-S/EdgeConv, 3) \textbf{b2}: ViG-S/GIN, and 4) \textbf{b3}: ViG-S/GraphSage. For the scalability analysis of the IOE, we also consider the PyramidViG-M as the alternative ViG backbone that sustains graph features dimensional reductions as the network deepens. We implemented PyramidViG-M to follow the feature dimensional reductions across stages as in \cite{han2022vision} and fixed four (04) blocks within each superblock in the supernet (recall \ref{subsubsec:supernet}).
     
\noindent{\large \textcircled{\small 2}} \textbf{HW-mapping baselines:} We consider the default \textit{standalone} deployment options – i.e., the full mapping of an entire ViG model to a singular CU (e.g., to the GPU only). We also consider hybrid mapping strategies in which inter-CU transitions are limited, as proposed in \cite{dagli2022axonn}.  

\noindent{\large \textcircled{\small 3}} \textbf{MAESTRO GNN baseline:} We use the aforementioned PyramidViG-M GIN-variant for our hardware scalability experiments using the MAESTRO simulator. For the convenience of MAESTRO, we define the GIN operation by its low-level implementations of the \textit{aggregation} and \textit{combination} phases. That is, the \textit{aggregation} entails a matrix multiplication between the adjacency matrix and the feature embedding matrix, whereas the \textit{combination} entails another matrix multiplication to transform the aggregated graph features to another representation for the following layer. 

\begin{figure}[t]
\centering
    \includegraphics[width=\textwidth]{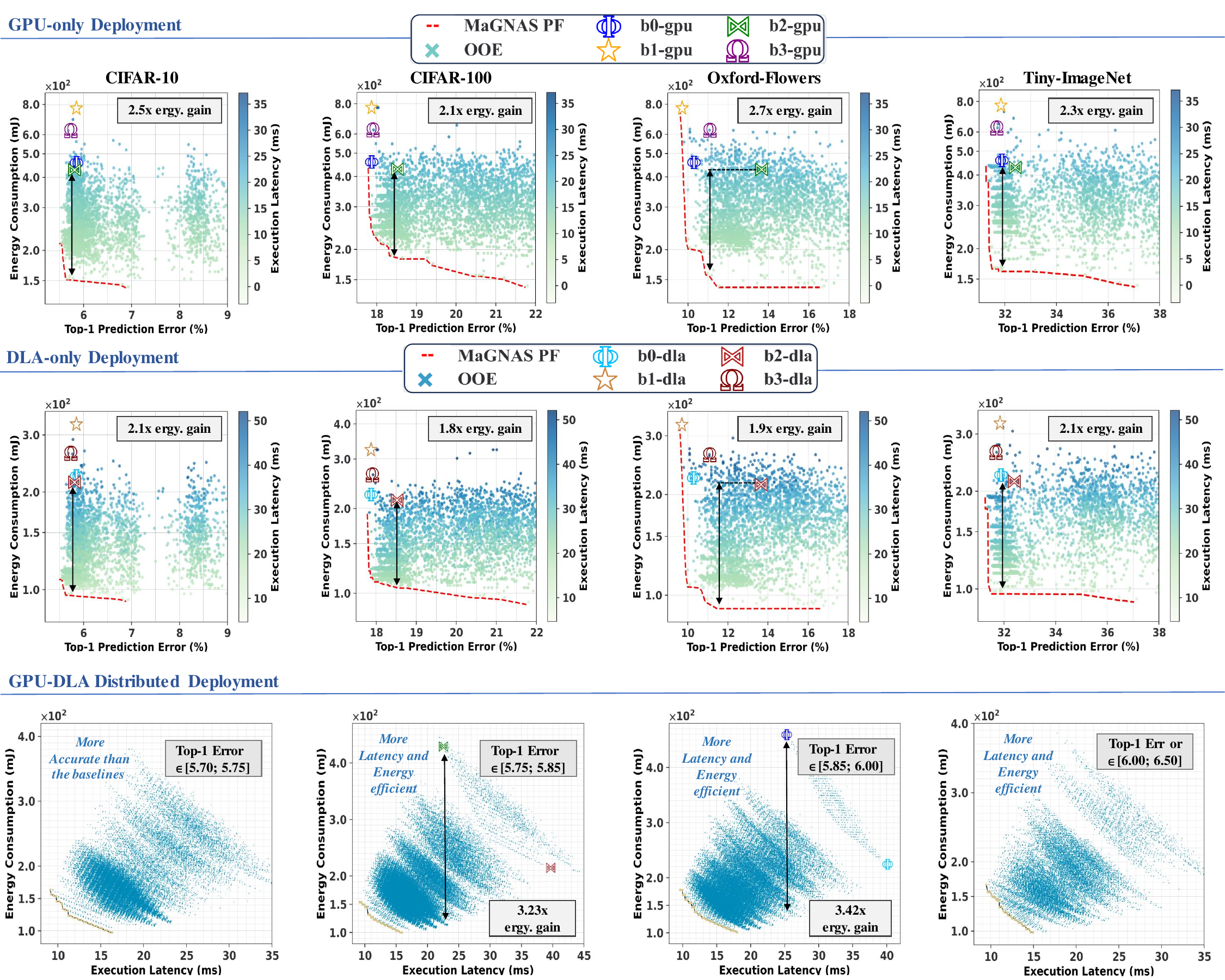}
    \caption{The first two rows show the performance of the explored GNNs in $(\mathbb{A})$ by the OOE on four datasets (\textit{from left to right}: a) CIFAR-10, b) CIFAR-100, c) Oxford-Flowers, and d) Tiny-ImageNet. The Hardware metrics  (i.e., latency and energy) are shown for \textit{GPU-only} deployment in the first row and for \textit{DLA-only} deployment in the second row. The third row shows the IOE results on CIFAR-10 grouped by prediction error intervals.} 
    \label{fig:paretos}
    \vspace{-2ex}
\end{figure}

\subsection{OOE Results: GNN Architecture Optimization}
We first examine the merit of the OOE in identifying GNN architectures that can achieve favorable performance trade-offs compared to the baselines.
In Figure \ref{fig:paretos}, the first two rows depict the explored GNN architectures from $\mathbb{A}$ by the OOE on the four (04) datasets given \textit{standalone} mapping strategies on GPU-only (\textit{top row}) and DLA-only (\textit{middle row}).
Compared to the baselines defined above, our obtained Pareto-optimal GNN architectures generally dominate all baselines on the four image classification datasets with regard to the three performance metrics of accuracy, latency, and energy consumption. Specifically, the OOE can identify GNN architectures that achieve up to $\sim$\textbf{3.6}$\times$ latency speedup than baselines when deployed onto the GPU; can realize up to $\sim$\textbf{2.8}$\times$ more energy efficiency gains compared to the baselines when deployed onto the DLA -- all while maintaining comparable accuracy scores. As will be emphasized in the subsequent Section \ref{Sec:Pareto_analysis}, the reasons for this dominance by the OOE's GNN architectures is attributed to the allowed diversification of \textit{Graph Op} across the different ViG superblocks (as specified in $\mathbb{A}$ from Table \ref{tab:joint_search_space}), which has facilitated achieving better accuracy-performance trade-offs.  
Moreover, skipping the \textit{FFN} and the \textit{Grapher's} FC pre-processing layers offers attractive design choices to avoid unnecessary computation, especially when the set of features is limited and can be already captured by the basic layers of the \textit{Grapher} modules -- which is the case for the simpler datasets (e.g., CIFAR-10). Our OOE recognized this property and leveraged its knowledge to concentrate its search on identifying GNN architectural parameters that achieve the best accuracy levels with the minimal number of \textit{FFN} and FC pre-processing layers. 

\subsection{IOE Results: Hardware Mapping Optimization}
\label{subsec:ioe_discussion}
We further assess the efficacy of the IOE in identifying effective mapping configurations for provided GNN architectures. 
The bottom row of Figure \ref{fig:paretos} shows the optimization results when exploring mapping strategies from $\mathbb{M}$ for the \textit{top-performing} GNN architectures (as ranked by equation \ref{eqn:OOE_rank}) provided to the IOE. The results are reported for CIFAR-10 and grouped by TOP-1 error intervals in each sub-figure. A similar trend has also been observed in the other datasets.  
At each top-1 error interval, we can observe that the IOE explored various mapping strategies, as illustrated by the latency-energy trade-offs. The bulk of these trade-offs are captured within the range of performance values from the standalone deployment options, that is, between the GPU-only and DLA-only mapping options' latency/energy consumption values, as depicted by the middle sub-figures. 
Remarkably, the explored configurations form distinguishable contours, each showing a specific GNN architecture alongside its explored mapping options -- represented by the different latency-energy trade-off values. Specifically, the GPU-only and DLA-only mapping configurations for each GNN architecture are located at the boundaries of its curved line. The intermediate points between the extremes illustrate the performance of the distributed deployment settings and show how each mapping configuration results in different latency-energy trade-offs.  

Furthermore, as both \textit{GNNs} and \textit{mappings} are considered together in the IOE design space, superior energy gains can be realized through more compact GNN architectures. 
For instance, as illustrated in the third sub-Figure, an energy gain up to \textbf{$\sim$3.42$\times$} can be attained compared to the \textbf{b2-gpu} while preserving comparable latency and accuracy levels by opting for another GNN architecture and distributed mapping. Upon comparing the curve lines, we can observe that GNN architectures that outperformed the baselines in the OOE (i.e., in the standalone deployment options shown by the extremes) typically maintain their dominance within the IOE and proves that rank is preserved across GNN architectures and mapping schemes in this joint search space. 

\begin{table}[ht]
\centering
\caption{Detailed performance results, GNN architectural parameters, and mapping strategies of our Pareto optimal models (\textbf{a0-a3}). The original ViG-S and its variants (\textbf{b0-b3}) on the four datasets on the NVIDIA Jetson Xavier AGX SoC platform. '\textbf{G}' and '\textbf{D}' in the latency and energy columns indicate GPU and DLA, respectively.}
\vspace{-1ex}
\fontsize{9}{9}\selectfont
\scalebox{0.67}
{
\label{tab:models_details}
\begin{tblr}{
  cells = {c},
  cell{2}{1} = {r=4}{},
  cell{6}{1} = {r=4}{},
  cell{10}{1} = {r=4}{},
  cell{14}{1} = {r=4}{},
  cell{18}{1} = {r=4}{},
  vline{3,7} = {-}{},
  hline{1-2,6,14,18,22} = {-}{},
  hline{10} = {1,7-10}{},
  hline{10} = {2-6}{0.03em},
}
\textbf{Datasets}                          & \textbf{GNN Models} & {\textbf{TOP-1 Acc}\\\textbf{(\%)}}                                                                                                                                                                                                                                                                                                            & {\textbf{Graph-Ops}\\\textbf{(M, E, G, S)}} & {\textbf{FFN-use }\\\textbf{(\%)}} & {\textbf{FC pre-use }\\\textbf{(\%)}} & {\textbf{Latency}\\\textbf{(ms)}}          & {\textbf{Energy}\\\textbf{(mJ)}}              & {\textbf{GPU-use}\\\textbf{(\%)}} & {\textbf{DLA-use }\\\textbf{(\%)}} \\
All-datasets                      & {\color{blue}$\Phi$}~Baseline-b0               & {\textbf{C10:} 94.15,~\textbf{C100:}~82.13\\\textbf{F:}~89.71,~\textbf{Ti:} 68.12}                                                                                                                                                                                                                                                             & M-M-M-M                                     & 100                                & 100                                    & {\textbf{G:} 25.28\\\textbf{D:}~40.11} & {\textbf{G:} 459.44\\\textbf{D:}~224.41}  & -                                 & -                                  \\
                                  & {\color{orange}$\bigstar$} Baseline-b1               & {\textbf{C10:}~94.15~\textbf{C100:}~82.13\\\textbf{F:}~90.29,~\textbf{Ti:}~68.15}                                                                                                                                                                                                                          & E-E-E-E                                     & 100                                & 100                                    & {\textbf{G:} 33.74\\\textbf{D:}~62.11} & {\textbf{G:} 770.36\\\textbf{D:}~ 323.70} & -                                 & -                                  \\
                                  & {\color{green}$\bowtie$}~Baseline-b2               & {\textbf{C10:}~94.20,~\textbf{C100:}~81.49\\\textbf{F:}~86.37,~\textbf{Ti:}~67.62}                                                                                                                                                 & G-G-G-G                                     & 100                                & 100                                    & {\textbf{G:} 22.49\\\textbf{D:}~39.62} & {\textbf{G:} 429.07\\\textbf{D:}~214.35} & -                                 & -                                  \\
                                  & {\color{brown}$\Omega$}~Baseline-b3               & {\textbf{C10:}~94.27,~\textbf{C100:}~82.10\\\textbf{F:}~88.92,~\textbf{Ti:}~68.32} & S-S-S-S                                     & 100                                & 100                                    & {\textbf{G:}~29.57\\\textbf{D:}~57.77} & {\textbf{G:} 623.76\\\textbf{D:}~263.48}  & -                                 & -                                  \\
{CIFAR-10 \\\textbf{(C10)}}       & {\color{red}$\bigcirc$} Ours-a0                   & 94.25                                                                                                                                                                                                                                                                                                                                          & G-G-G-G                                     & 25                                 & 25                                     & 16.02                                      & 97.0                                          & 09                                & 91                                 \\
                                  & {\color{red}$\bigcirc$} Ours-a1                   & \textbf{94.46}                                                                                                                                                                                                                                                                                                                                 & G-G-G-G                                     & 100                                & 0                                      & 19.49                                      & 118.00                                        & 17                                & 83                                 \\
                                  & {\color{red}$\bigcirc$} Ours-a2                   & 94.32                                                                                                                                                                                                                                                                                                                                          & G-M-G-G                                     & 25                                 & 0                                      & \textbf{11.19}                             & 121.14                                        & 75                                & 25                                 \\
                                  & {\color{red}$\bigcirc$} Ours-a3                   & 94.32                                                                                                                                                                                                                                                                                                                                          & G-M-G-G                                     & 25                                 & 0                                      & 14.18                                      & \textbf{105.11}                               & 33                                & 67                                 \\
{CIFAR-100\\\textbf{(C100)}}      & {\color{red}$\bigcirc$} Ours-a0                   & 82.13                                                                                                                                                                                                                                                                                                                                          & S-G-S-G                                     & 100                                & 25                                     & 17.72                                      & 180.56                                        & 50                                & 50                                 \\
                                  & {\color{red}$\bigcirc$} Ours-a1                   & \textbf{82.17}                                                                                                                                                                                                                                                                                                                                 & S-S-S-S                                     & 100                                & 75                                     & 34.72                                      & 271.62                                        & 30                                & 70                                 \\
                                  & {\color{red}$\bigcirc$} Ours-a2                   & 81.63                                                                                                                                                                                                                                                                                                                                          & G-G-G-G                                     & 50                                 & 50                                     & \textbf{15.06}                             & \textbf{131.81}                               & 50                                & 50                                 \\
                                  & {\color{red}$\bigcirc$} Ours-a3                   & 82.13                                                                                                                                                                                                                                                                                                                                          & S-G-S-G                                     & 100                                & 25                                     & 17.29                                      & 197.80                                        & 55                                & 45                                 \\
{Oxford-Flowers\\\textbf{(F)}}    & {\color{red}$\bigcirc$} Ours-a0                   & \textbf{89.90}                                                                                                                                                                                                                                                                                                                                 & M-G-M-M                                     & 75                                 & 75                                     & 14.37                                      & 153.54                                        & 69                                & 31                                 \\
                                  & {\color{red}$\bigcirc$} Ours-a1                   & 88.43                                                                                                                                                                                                                                                                                                                                          & G-G-G-G                                     & 0                                  & 50                                     & \textbf{9.60}                              & 119.07                                        & 90                                & 10                                 \\
                                  & {\color{red}$\bigcirc$} Ours-a2                   & 88.43                                                                                                                                                                                                                                                                                                                                          & G-G-G-G                                     & 0                                  & 50                                     & 12.30                                      & \textbf{105.88}                               & 40                                & 60                                 \\
                                  & {\color{red}$\bigcirc$} Ours-a3                   & 89.02                                                                                                                                                                                                                                                                                                                                          & M-G-G-G                                     & 25                                 & 25                                     & 12.82                                      & 116.63                                        & 50                                & 50                                 \\
{Tiny-ImageNet   \\\textbf{(Ti)}} & {\color{red}$\bigcirc$} Ours-a0                   & 68.40                                                                                                                                                                                                                                                                                                                                          & M-G-G-G                                     & 25                                 & 0                                      & \textbf{13.07}                             & 114.89                                        & 50                                & 50                                 \\
                                  & {\color{red}$\bigcirc$} Ours-a1                   & 68.40                                                                                                                                                                                                                                                                                                                                          & M-G-G-G                                     & 25                                 & 0                                      & 15.47                                      & \textbf{102.06}                               & 17                                & 83                                 \\
                                  & {\color{red}$\bigcirc$} Ours-a2                   & \textbf{68.51}                                                                                                                                                                                                                                                                                                                                 & M-G-G-G                                     & 75                                 & 25                                     & 16.37                                      & 122.56                                        & 38                                & 62                                 \\
                                  & {\color{red}$\bigcirc$} Ours-a3                   & \textbf{68.51}                                                                                                                                                                                                                                                                                                                                 & M-G-G-G                                     & 75                                 & 25                                     & 17.87                                      & 115.78                                        & 19                                & 81                                 
\end{tblr}
}
\end{table}

\subsection{Analysis of Pareto Search and Models} \label{Sec:Pareto_analysis}

\subsubsection{Results Discussion}
In Table \ref{tab:models_details}, we provide a detailed analysis of performances, architectural parameters, and mapping strategies of the ViG baselines \textbf{[b0-b3]} and a selection of our final Pareto optimal models from the two-tier search \textbf{[a0-a3]} for each dataset. 
As shown, although our models maintain comparable accuracy scores to the baselines, they generally achieve better speedups and energy efficiency results. To be more precise, our models achieve on average $\sim$\textbf{1.57}$\times$ and $\sim$\textbf{2.49}$\times$ latency speedups; $\sim$\textbf{3.38}$\times$ and $\sim$\textbf{1.65}$\times$ more energy efficiency when compared against the original ViG baseline fully-deployed onto the GPU and DLA, respectively. This dominance is primarily attributed to 3 factors: (\emph{i}) the enabled diversification of Graph Op parameter throughout the ViG superblocks, which enables interleaving both \textit{powerful} and \textit{resource-efficient} operators within a model architecture. For instance, examining the Oxford-Flowers results in the Table, model \textbf{a0} interleaves both $\mathtt{Max\text{-}Relative}$ and $\mathtt{GIN}$ operators. 
The former contributes to the model's representational capacity and compensates for the inadequacy of $\mathtt{GIN}$ operators in capturing long-range dependencies from the graph nodes features, ultimately leading the model to surpass baseline \textbf{b0}'s accuracy score (89.9\% to 89.71\%). On the other hand, the employment of $\mathtt{GIN}$ operator -- alongside other factors -- leads \textbf{a0} to achieve superior latency and energy efficiency scores. (\emph{ii}) The additional varying architectural parameters from $\mathbb{A}$ (e.g., FFN-use) enable tuning the model's size and learning capacity to the task and dataset complexity. (\emph{iii}) The distributed mapping strategies, as indicated by the 
\textit{GPU-use} and \textit{DLA-use} columns in Table \ref{tab:models_details}, further balance the latency-energy trade-offs by effectively utilizing different CUs. 

\begin{figure}[ht]
\centering
\includegraphics[width=\textwidth]{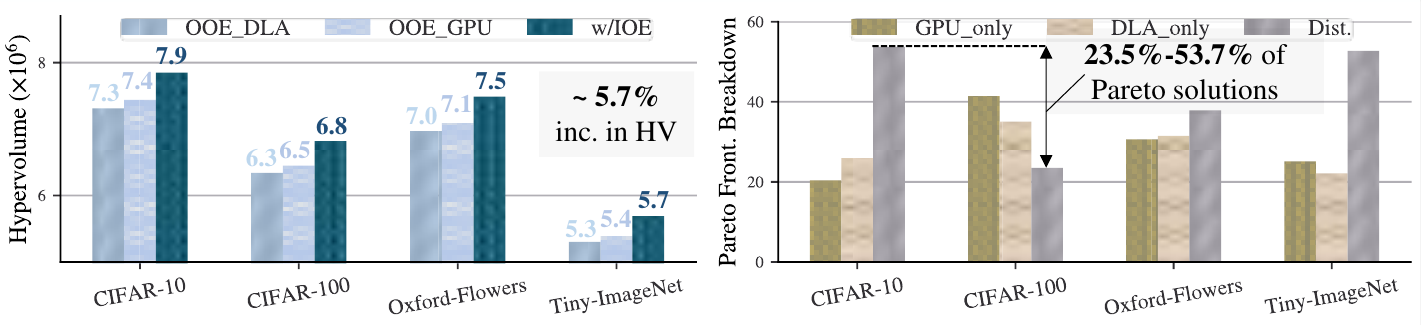}
\vspace{-5ex}
    \caption{\textit{Left}: Hypervolume analysis when including the IOE against those of the standalone OOE for the DLA and GPU. \textit{Right}: Breakdown of the combined Pareto Fronts constituents on the basis of mapping options.}
    \label{fig:HV_PF}
    \vspace{-2ex}
\end{figure}

\subsubsection{Hypervolume and Pareto Composition Analysis}
To appraise the efficiency of our nested evolutionary search algorithm in identifying meaningful and mapping configurations, we compare its Hypervolume \cite{shang2020survey} against those of baseline OOE searches conducted on the \textit{standalone} deployment options on the GPU and DLA. Succinctly, the Hypervolume measures the volume of the dominated area in the objective space by the estimated Pareto fronts. In Figure \ref{fig:HV_PF} (\textit{left}), we can observe that the nested search (\textit{w/IOE}) improves the Hypervolume scores over the baseline $OOE\_GPU$ search by $\sim$\textbf{5.7}\% on average across the four (04) datasets, indicating the IOE's merit in extending the dominated area in the search space. In Figure \ref{fig:HV_PF} (\textit{right}), we complement the Hypervolume analysis with a breakdown of the Pareto front composition with regard to the mapping strategies. Specifically, we consider the non-dominated solutions by combining Pareto fronts obtained at every generation. As seen, the \textit{distributed} mapping options constitute 23.5\%-53.7\% of the solutions on the Pareto front, indicating their value in elevating resource efficiency for the various models.

\subsubsection{Analysis of GNN workload distribution}
\label{subsec:mapping_details}

In this subsection, we showcase how different GNN workload assignments across the GPU and DLA influence the latency-energy tradeoffs. In Table \ref{tab:gnn_workloads}, we select one of the Pareto-optimal models, \textbf{Ours-a3} on CIFAR-100, and compare three mapping configurations: (\emph{i}) Standalone options in which the model is fully deployed on either GPU or DLA. (\emph{ii}) Constrained transition options (as introduced in \cite{dagli2022axonn}) where the number of allowable inter-CU transitions is limited to those that offer the best tradeoffs in order to mitigate data transmission overheads (i.e., the write-back and initial cold cache misses). (\emph{iii}) Ours (IOE) are the mapping options provided through our IOE with unconstrained inter-CU transitions. 

\begin{table}[ht]
\centering
\caption{Details and comparison of the GNN workload Assignment. `G' and `D' indicate GPU and DLA assignment, respectively. Note that each Grapher block is first succeeded by a corresponding FFN block.}
\vspace{-1ex}
\fontsize{9}{9}\selectfont
\scalebox{0.95}
{
\label{tab:gnn_workloads}
\begin{tabular}{|l|c|c|c|c|c|c|c|}
\hline
\multicolumn{1}{|c|}{\textbf{Mapping option}} & \textbf{Stem} & \textbf{Grapher} & \textbf{FFN} & \textbf{Cls} & \textbf{\#transit} & \textbf{Lat.} & \textbf{Enrg.} \\ \hline
DLA-only        & {\color{brown}D} & {\color{brown}D}-{\color{brown}D}-{\color{brown}D}-{\color{brown}D}-{\color{brown}D}-{\color{brown}D}-{\color{brown}D}-{\color{brown}D} & {\color{brown}D}-{\color{brown}D}-{\color{brown}D}-{\color{brown}D}-{\color{brown}D}-{\color{brown}D}-{\color{brown}D}-{\color{brown}D} & {\color{brown}D} & 0  & 25.56          & \textbf{121.74} \\
GPU-only        & {\color{darkgreen}G} & {\color{darkgreen}G}-{\color{darkgreen}G}-{\color{darkgreen}G}-{\color{darkgreen}G}-{\color{darkgreen}G}-{\color{darkgreen}G}-{\color{darkgreen}G}-{\color{darkgreen}G} & {\color{darkgreen}G}-{\color{darkgreen}G}-{\color{darkgreen}G}-{\color{darkgreen}G}-{\color{darkgreen}G}-{\color{darkgreen}G}-{\color{darkgreen}G}-{\color{darkgreen}G} & {\color{darkgreen}G} & 0  & \textbf{13.42} & 273.22          \\ \hline
constr-transit1 & {\color{brown}D} & {\color{brown}D}-{\color{darkgreen}G}-{\color{darkgreen}G}-{\color{darkgreen}G}-{\color{darkgreen}G}-{\color{darkgreen}G}-{\color{darkgreen}G}-{\color{darkgreen}G} & {\color{brown}D}-{\color{darkgreen}G}-{\color{darkgreen}G}-{\color{darkgreen}G}-{\color{darkgreen}G}-{\color{darkgreen}G}-{\color{darkgreen}G}-{\color{darkgreen}G} & {\color{darkgreen}G} & 1  & 16.31 & 232.60          \\
constr-transit1 & {\color{darkgreen}G} & {\color{darkgreen}G}-{\color{darkgreen}G}-{\color{darkgreen}G}-{\color{darkgreen}G}-{\color{darkgreen}G}-{\color{brown}D}-{\color{brown}D}-{\color{brown}D} & {\color{darkgreen}G}-{\color{darkgreen}G}-{\color{darkgreen}G}-{\color{darkgreen}G}-{\color{darkgreen}G}-{\color{brown}D}-{\color{brown}D}-{\color{brown}D} & {\color{brown}D} & 1  & 17.42          & 226.79          \\ \hline
constr-transit2 & {\color{brown}D} & {\color{brown}D}-{\color{darkgreen}G}-{\color{darkgreen}G}-{\color{darkgreen}G}-{\color{darkgreen}G}-{\color{darkgreen}G}-{\color{darkgreen}G}-{\color{brown}D} & {\color{brown}D}-{\color{darkgreen}G}-{\color{darkgreen}G}-{\color{darkgreen}G}-{\color{darkgreen}G}-{\color{darkgreen}G}-{\color{darkgreen}G}-{\color{brown}D} & {\color{brown}D} & 2  & 17.58          & 220.23          \\
constr-transit2 & {\color{darkgreen}G} & {\color{darkgreen}G}-{\color{darkgreen}G}-{\color{brown}D}-{\color{brown}D}-{\color{brown}D}-{\color{darkgreen}G}-{\color{darkgreen}G}-{\color{darkgreen}G} & {\color{darkgreen}G}-{\color{darkgreen}G}-{\color{brown}D}-{\color{brown}D}-{\color{brown}D}-{\color{darkgreen}G}-{\color{darkgreen}G}-{\color{darkgreen}G} & {\color{darkgreen}G} & 2  & 17.11 & 227.15          \\ \hline
Ours (IOE)        & {\color{brown}D} & {\color{darkgreen}G}-{\color{darkgreen}G}-{\color{darkgreen}G}-{\color{darkgreen}G}-{\color{darkgreen}G}-{\color{darkgreen}G}-{\color{darkgreen}G}-{\color{darkgreen}G} & {\color{darkgreen}G}-{\color{brown}D}-{\color{brown}D}-{\color{brown}D}-{\color{brown}D}-{\color{darkgreen}G}-{\color{brown}D}-{\color{brown}D} & {\color{brown}D} & 12 & \textbf{17.29} & \textbf{197.8}  \\ \hline
\end{tabular}
}
\vspace{-1ex}
\end{table}

As no single optimal solution exists for any distributed mapping strategy, we ensure a fair comparison between our approach and the constrained transition strategies by comparing evaluations of one objective function (energy) while fixing the other (latency). As such, for each constrained transition option, we use two (02) Pareto optimal solutions whose latency values are closest to our solution -- i.e., solutions with latency closest to 17.29 ms. From the reported results in Table \ref{tab:gnn_workloads}, we can observe that with our unconstrained mapping strategy, a single inference sustains 197.8 mJ on average, which is more efficient than the best energy numbers, 226.79 mJ and 220.23 mJ, experienced by each of the other distributed mapping baselines, `constr-transit1' and `constr-transit2', respectively. The reasons for this improvement can be attributed to the following: (\emph{i}) graph feature sizes are relatively small throughout the ViG models, leading to low inter-CU transmission overhead penalties to be experienced on the Xavier SoC. As Such, our IOE optimization strategy was able to exploit this property to identify more efficient mapping configurations with a larger number of transitions. (\emph{ii}) Each computing block type within the ViG exhibits different affinities towards the underlying CUs. Thus, our IOE optimization strategy leveraged the other property of unconstrained transitions to map as many Grapher blocks to the GPU as feasible and as many FFN blocks to the DLA as possible before transmission costs become non-negligible. 

\begin{figure}[ht]
\centering
    \includegraphics[width=\textwidth]{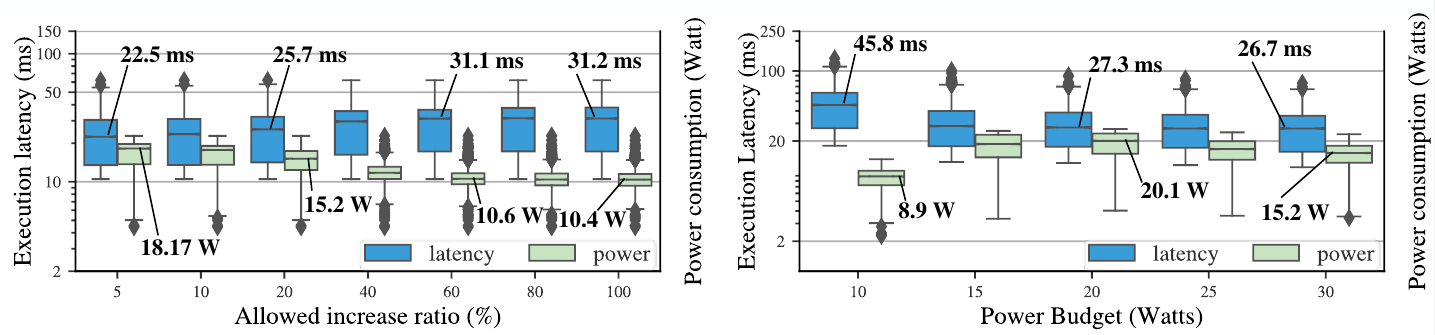}
    \caption{Results of the two constrained optimization: Latency and power consumption numbers are reported under variation of (\textit{Left}) the allowable latency increase ratio compared to GPU-only, and (\textit{right}) the available power budget. Numbers indicate median values.}
    \label{fig:constr_search}
\end{figure}

\subsection{Constraint-aware Optimization}
\label{subsec:constrained_optim}
As many embedded systems employ real-time execution requirements, we test the effectiveness of our framework when the search algorithm is performed under strict latency and power constraints. In particular, we specify two experiments, each associated with one of the following constraints: \textbf{(1) Latency}, in which the constraint specifies the max allowable increase in latency compared to the \textit{standalone} deployment option on the fastest SoC component (i.e., GPU-only). \textbf{(2) Power budget}; by fixing low values of clock frequencies and a limited number of CPU cores and memory speed transmission \cite{power_jetson}. The first constraint is common for real-time systems governed by strict execution deadlines, whereas the second constraint is more common for battery-powered systems operating on limited power budgets.
We conduct the two constrained optimization on the IOE using baselines [\textbf{b0-b3}] and our models [\textbf{a0-a3}] on the CIFAR-100. 
\begin{wraptable}{r}{0.45\textwidth}
  \centering
  \vspace{-5mm}
  \caption{Workload distribution.}
  \vspace{-4mm}
  \label{tab:lat_const_distr}
  \fontsize{9}{9}\selectfont
    \scalebox{0.75}
    {
   \begin{tabular}{|c|rrrrrrr|}
    \hline
    \multirow{2}{*}{\textbf{\begin{tabular}[c]{@{}c@{}}Workload \\ Distribution\end{tabular}}} & \multicolumn{7}{c|}{\textbf{Allowable latency increase ratio (\%)}} \\ \cline{2-8} 
     & \multicolumn{1}{r|}{5} & \multicolumn{1}{r|}{10} & \multicolumn{1}{r|}{20} & \multicolumn{1}{r|}{40} & \multicolumn{1}{r|}{60} & \multicolumn{1}{r|}{80} & 100 \\ \hline
    \textbf{\begin{tabular}[c]{@{}c@{}}Avg. GPU\\ utilization\end{tabular}} & \multicolumn{1}{r|}{0.97} & \multicolumn{1}{r|}{0.91} & \multicolumn{1}{r|}{0.74} & \multicolumn{1}{r|}{0.56} & \multicolumn{1}{r|}{0.50} & \multicolumn{1}{r|}{0.50} & 0.50 \\ \hline
    \textbf{\begin{tabular}[c]{@{}c@{}}Avg. DLA\\ utilization\end{tabular}} & \multicolumn{1}{r|}{0.03} & \multicolumn{1}{r|}{0.09} & \multicolumn{1}{r|}{0.26} & \multicolumn{1}{r|}{0.44} & \multicolumn{1}{r|}{0.50} & \multicolumn{1}{r|}{0.50} & 0.50 \\ \hline
    \end{tabular}
    }
  \vspace{-4mm}
\end{wraptable}
We report the absolute latency and average power consumption values in Figure \ref{fig:constr_search}. 
We also characterize the latency constraint by enforcing a max \textit{allowable increase ratio} from the fastest CU (i.e., the GPU). As shown in left Figure \ref{fig:constr_search}, low latency increase ratio ($\le 20\%$) leads the IOE towards delegating more computation kernels to the GPU, resulting in more power-demanding mapping strategies. Compared to the soft-constraint case (i.e.,w/ tolerance of 100\% increase in latency), the power demands at an allowed increase ratio of 5\% are 1.75$\times$ more. 

As the tolerable increase ratio rises ($\ge 30\%$), the constraint on the search is gradually relaxed. As shown in Table \ref{tab:lat_const_distr}, the optimizer gains more freedom in exploring mapping options and favors delegating more computation kernels to the DLA for energy efficiency. The power efficiency gains start to plateau around a 50\% increase ratio, indicating that the IOE has converged onto mapping strategies that maximize the fitness formula (as defined in  (\ref{eqn:ioe_fitness})) by balancing latency and power efficiency. This convergence is sensible given how between the GPU and DLA, one component is roughly twice as effective as the other with regards to one performance objective, i.e., execution latency on the GPU is almost 2$\times$ less than the DLA, and the DLA incurs 2$\times$ less power consumption than the GPU (see Table \ref{tab:models_details}); given that we assigned equivalent weights for the objectives in the fitness evaluation formula in (\ref{eqn:ioe_fitness}), i.e.,  $\gamma_1 = \gamma_2 = 1$.  
\begin{wraptable}{r}{0.35\textwidth}
  \centering
  \vspace{-6mm}
  \hspace{-4mm}
  \caption{Workload distribution.}
  \vspace{-4mm}
  \label{tab:pwr_const_distr}
  \fontsize{9}{9}\selectfont
    \scalebox{0.72}
    {
   \begin{tabular}{|c|rrrrr|}
    \hline
    \multirow{2}{*}{\textbf{\begin{tabular}[c]{@{}c@{}}Workload \\ Distribution\end{tabular}}} & \multicolumn{5}{c|}{\textbf{Available Power Budget (mW)}} \\ \cline{2-6} 
     & \multicolumn{1}{r|}{10} & \multicolumn{1}{r|}{15} & \multicolumn{1}{r|}{20} & \multicolumn{1}{r|}{25} & 30 \\ \hline
    \textbf{\begin{tabular}[c]{@{}c@{}}Avg. GPU \\ utilization\end{tabular}} & \multicolumn{1}{r|}{0.74} & \multicolumn{1}{r|}{0.76} & \multicolumn{1}{r|}{0.88} & \multicolumn{1}{r|}{0.88} & 0.81 \\ \hline
    \textbf{\begin{tabular}[c]{@{}c@{}}Avg. DLA \\ utilization\end{tabular}} & \multicolumn{1}{r|}{0.26} & \multicolumn{1}{r|}{0.24} & \multicolumn{1}{r|}{0.13} & \multicolumn{1}{r|}{0.13} & 0.19 \\ \hline
    \end{tabular}
    }
  \vspace{-4mm}
\end{wraptable}
The second experiment depicted in the right Figure \ref{fig:constr_search} shows that at tighter power budget constraints, the IOE focuses on identifying power-efficient mapping options at the expense of a slight decrease in latency, resulting in mappings that assign more GNN workloads to the DLA as depicted in Table \ref{tab:pwr_const_distr}. 
We note that in this experiment, we also maintain the latency minimization as objective, which also explains the low DLA utilization ratio values reported in Table \ref{tab:pwr_const_distr}. For instance, to satisfy the 10 Watts power constraint, the IOE specifies mapping settings with a median latency of 45.8\% -- 1.71$\times$ more than the latency experienced at a power budget of 30 Watts. More latency-efficient mappings are identified with refined workload distribution as the power budgets are relaxed. 

\begin{figure}[tp]
\centering
    \includegraphics[width=\textwidth]{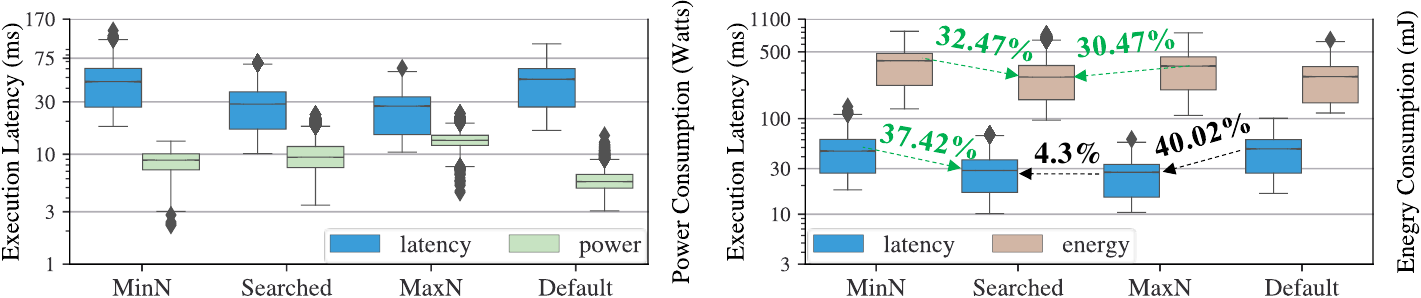}
    \caption{Ablation on the impact of including DVFS optimization within the IOE. \textbf{Searched} DVFS is compared against the \textbf{MinN}, \textbf{MaxN}, and \textbf{Default} settings with regards to 
    (\textit{Left)}: Latency-Power trade-offs, and (\textit{Right}): Latency-Energy trade-offs. Numbers in the right Figure indicate percentage change in values.}
    \label{fig:dvfs_ablation}
    \vspace{-1ex}
\end{figure}

\subsection{Ablation study on the impact of DVFS}
\label{subsec:dvfs_discussion}

In this experiment, we assess the merit of including DVFS optimization within the IOE. We reuse the baselines [b0-b3] and our models [a0-a3] from the CIFAR-100 experiment. Their mappings are kept fixed, and we run the models through the DVFS optimization engine to assess how performance can be further enhanced. Specifically, we consider the following DVFS settings: (\emph{i}) \textbf{MaxN}, which resembles the high-performance mode on the Jetson Xavier SoC with clock frequencies set to the maximum. (\emph{ii}) \textbf{MinN}, which is an opposing best-effort mode for low-power operation in which clock frequencies are set to the minimum. (\emph{iii}) \textbf{Searched}; in which DVFS settings are searchable within the IOE (see Table \ref{tab:joint_search_space} for the values). \emph{iv}) \textbf{Default}; in which we use the default dynamic DVFS heuristic with CPU and GPU governors fixed to \textit{Schedutil}, \textit{nvhost podgov}, respectively. In this last setting, clock frequencies are dynamically adjusted at runtime depending on the underlying resources utilization, where clock frequencies are ranged from 0 to the maximum value on each component. We note that in addition to the GPU and DLA frequency variations, we also scale the CPU and EMC clock frequencies as both influence data transmissions between the shared system memory and private memories of GPU/DLA.
We run the IOE with the same optimization parameters to ensure a fair evaluation. In Figure \ref{fig:dvfs_ablation}, we illustrate the performance trade-offs as incurred by the explored (\textit{GNN architectures} $\times$ \textit{HW mappings}) under the 04 DVFS settings. As expected, the left subfigure shows that the Searched mode exhibits a balanced trade-off between latency and power compared to the MinN and MaxN modes. More interestingly, however, the \textit{Searched} setting is able to identify configurations that yield superior energy gains to the fixed DVFS modes. In particular, the median latency and energy consumption values of Searched are \textbf{37.42}\% and \textbf{32.47}\% \textit{less} than MinN, respectively. On the other hand, though Searched incurs a \textbf{4.3}\% increase in its median latency compared to MaxN, it can achieve an order of magnitude more energy savings reaching \textbf{30.47}\%. This implies that the IOE identified the DVFS as a viable tuning knob to enhance energy efficiency by scaling clock frequencies across the different components. 
Moreover, latency in Searched is improved by \textbf{40.02\%} compared to the default DVFS governor. This is explained by the underlying logic of the dynamic heuristic, which only considers the hardware utilization and overlooks workload properties such as computation and memory requirements. For instance, memory-bounded workloads may benefit from GPU/DLA core downscaling with reduced energy at the same latency level. These properties are captured in our Search mode as we adjust the frequencies according to the GNN and mapping configurations. In addition, The default governors are set to avoid the idle state when the computing unit is not used, by lowering the frequency to 0, which helps in minimizing the power consumption (as shown in the left subfigure) but also worsens the execution latency as computing units usually need a warm-up stage to operate steadily after swapping between low and high frequencies.

\subsection{Generality and Scalability} \label{subsec:generality}

Employing an evolutionary algorithm (EA) for the IOE may seem excessive when the backbone ViG architecture is an isotropic one that does not experience feature map sizes change and when the mapping is performed across merely 02 CUs. As such, we perform an additional set of experiments in which we assess the scalability and generality of the IOE on the search-space levels of: (\emph{i}) \textit{\textbf{the ViG architectural backbone}}; where the supernet's backbone is implemented as a pyramid variant that allows dimensional reductions from one superblock to the next (recall Figure \ref{fig:vig_supernet}), unlike the aforementioned isotropic counterpart, and (\emph{ii}) \textit{\textbf{the hardware CUs}}; by simulating a case with 03 heterogeneous CUs. The details are provided below.

\subsubsection{On the ViG architectural level}
Using the Nvidia Xavier SoC with GPU and DLA, we compare the mapping results from the IOE between the isotropic (ViG-S) and pyramid (PyramidViG-M) variants (recall Section \ref{subsubsec:base}). As we analyze the effectiveness of the inner EA, we fix the GNN from the OOE for both variants by setting the design parameters, $\mathbb{A}$, in Table 1 (i.e., d=4, \textit{Graph Op}=\texttt{GIN}, fc\_use=False, ffn\_use=False, w=192), and specify an optimization budget of 2$\times10^{4}$ evaluations. 

\begin{wrapfigure}{r}{0.58\textwidth}
  \centering
  \vspace{-5mm}
  \hspace{-3mm}
    \includegraphics[width=0.58\textwidth]{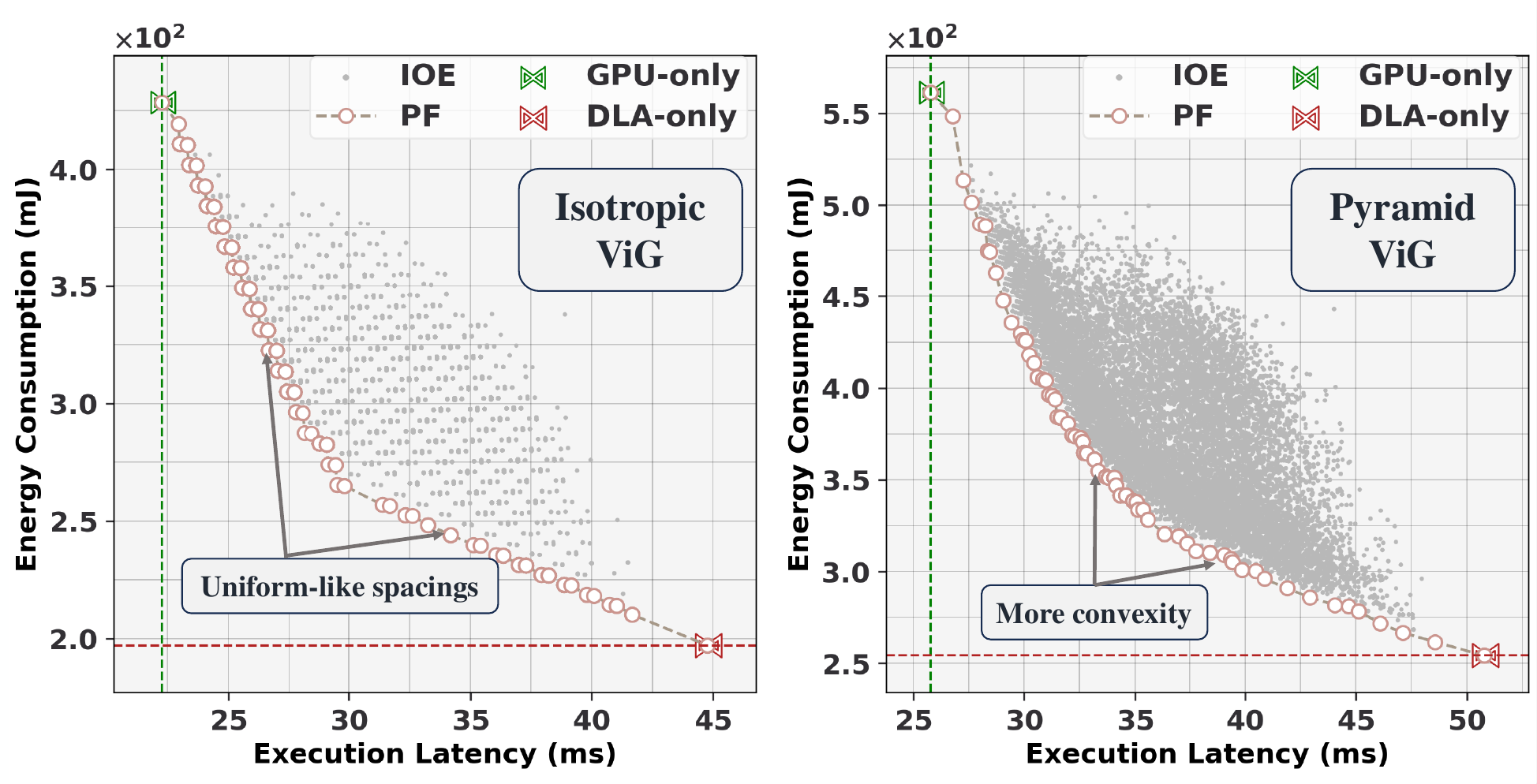}
  \vspace{-3.5mm}
  \caption{The results of the IOE EA optimization on the Isotropic Vision GNN (\textit{left}) and Pyramid Vision GNN (\textit{right}).}
  \vspace{-4.5mm}
  \label{fig:iso_pyramid}
\end{wrapfigure}

As depicted in Figure \ref{fig:iso_pyramid}, we can observe in the left subfigure that for the isotropic ViG, the explored mapping options follow well-defined spaced patterns between the two mapping extremes of \textit{GPU-only} and \textit{DLA-only}, offering almost uniform linear trade-offs between the energy efficiency and execution latency across various mapping options on the Pareto front. This results from the Grapher and FFN blocks being replicated throughout an isotropic architecture. 
As such, the performance evaluation of the different mapping options becomes predominantly influenced by the percentage of Grapher/FFN blocks assigned to each CU, irrespective of their order. Given such a setting, a scalarization method can be sufficient to determine the Pareto front by varying the ratio of mappable workloads on either CU. However, for the PyramidViG on the right, this property does not hold as each Grapher/FFN block entertains different dimensions of their input and output features depending on its position, leading to varying performance characterizations. As such, we observe that the sampled mapping options are more diverse in their energy and latency characterizations and that the Pareto front exhibits stronger convexity than its isotropic counterpart, reflecting a diverse, more complex mapping space.

\subsubsection{On the hardware CU level} 
Using the PyramidViG-M, we investigate how MaGNAS scales when the search space is further compounded with an increasing number of viable CUs. We simulate such use-case using MAESTRO tool \cite{kwon2020maestro} to specify 3 DSAs of diverse dataflows for CU heterogeneity (see the details in \ref{subsubsec:sim}). As every layer within MAESTRO is defined via low-level implementations (including \textit{aggregation} and \textit{combination} layers), we can characterize processing overheads within PyramidViG-M on a \textit{layerwise} basis and combine them to characterize larger blocks (e.g., Grapher).
At this point, we find that each `layer' rather than `block' can exhibit different performance characteristics at different ViG stages. For instance, the \textit{aggregation} sustains a substantial overhead when processing the sizable graph feature matrices at earlier blocks. This is predominantly due to the DSAs in MAESTRO not being implemented initially to support graph acceleration -- similar to how numerous SoC platforms (e.g., the Xavier) do not widely integrate specialized graph acceleration engines. As such, we can simulate an additional case to study the mapping on a \textit{layerwise} granularity to assess further how the EA in the IOE scales when the number of mappable options dramatically increase. To provide context, the mapping space of the PyramidViG-M is $\mathcal{O}$(1.72$\times$10$^{12}$) in the \textit{blockwise} using 2 CUs; $\mathcal{O}$(1.67$\times$10$^{16}$) in the \textit{blockwise} using 3 CUs; and $\mathcal{O}$(1.67$\times$10$^{23}$) in the \textit{layerwise} 3 CUs case, indicating an increasing level of problem complexity. 
\begin{wrapfigure}{r}{0.58\textwidth}
  \centering
  \vspace{-5mm}
  \hspace{-3mm}
    \includegraphics[width=0.58\textwidth]{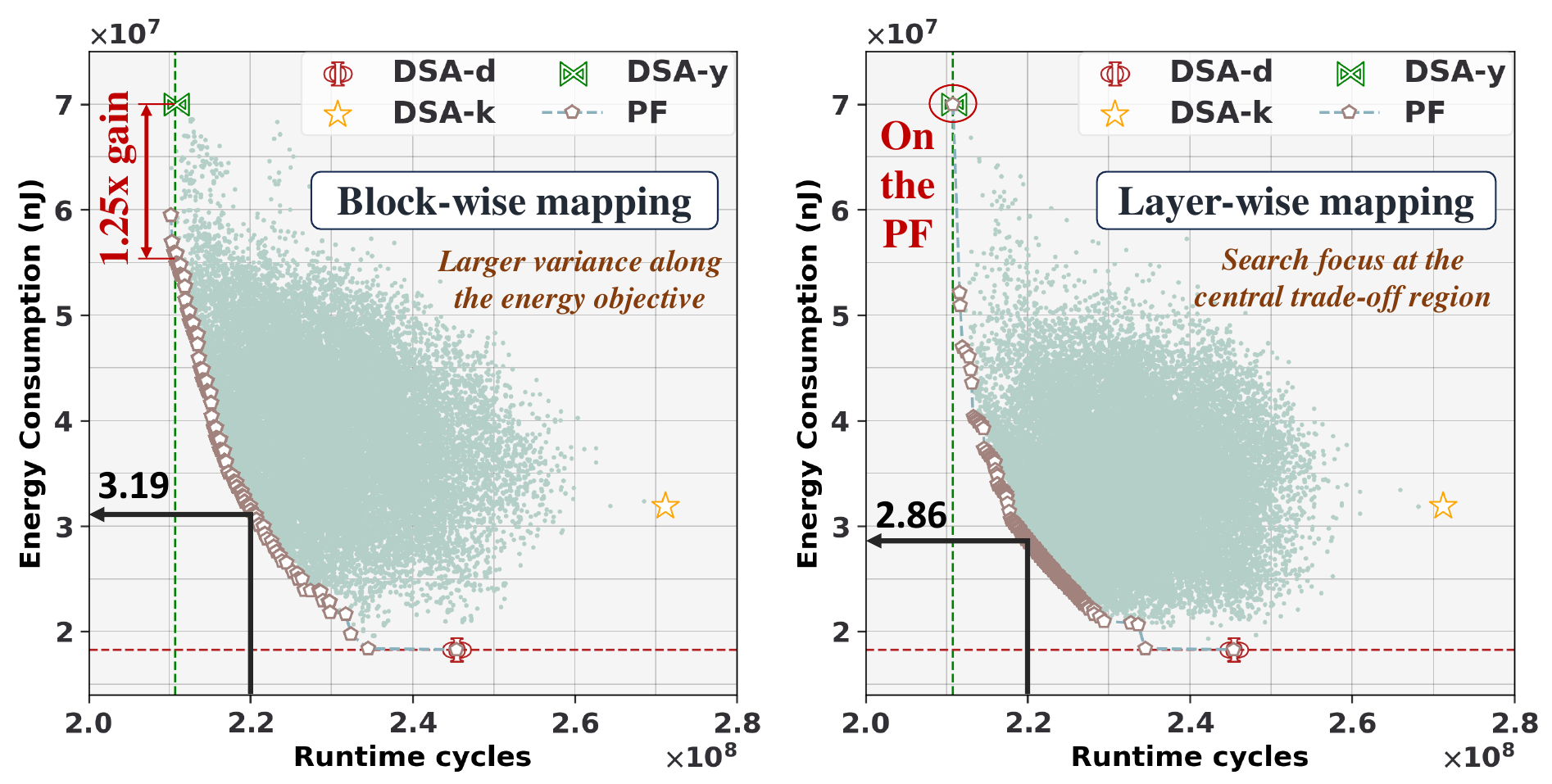}
  \vspace{-3.5mm}
  \caption{The results of the IOE optimization on MAESTRO \cite{kwon2020maestro} with: i) \textit{Block-wise mapping granularity} (\textit{left}) and ii) \textit{Layer-wise mapping granularity} (\textit{right}).}
  \vspace{-4.5mm}
  \label{fig:maestro_mapping}
\end{wrapfigure} 
In Figure \ref{fig:maestro_mapping}, we demonstrate how the inner EA scales effectively as the search space is expanded from the \textit{blockwise} to the \textit{layerwise} mapping granularity. We first specify a fixed optimization budget of 6$\times10^{4}$ evaluations for both. Moreover, although fully deploying the architecture on DSA-d completely dominates DSA-k deployment, the latter is still included since it represents the optimal mapping option for some individual layers. In the blockwise case (\textit{left}), we observe that the EA focuses on exploring more mapping solutions at the energy consumption extremes due to coarse-grained characterization of the Grapher block, leading it to identify distributed mapping options that dominate the standalone extreme, i.e., the EA identifies a distributed mapping configuration that achieves 1.25$\times$ energy gains over DSA-y for the same latency level. The opposite occurs for the \textit{layerwise} search, where despite the much larger optimization space, the EA was capable of recognizing benefits from distributing the \textit{aggregation} and \textit{combination} across different DSAs, leading it to concentrate the search more at the centralized latency-energy trade-off region. For example, at execution latency of $\sim2.2\times10^8$ cycles, the layerwise search by the IOE was able to identify a mapping option that incurs $28.6$ mJ compared to $31.9$ mJ from the blockwise search.

\begin{wrapfigure}{r}{0.42\textwidth}
  \centering
  \vspace{-4mm}
  \hspace{-3mm}
    \includegraphics[width=0.42\textwidth]{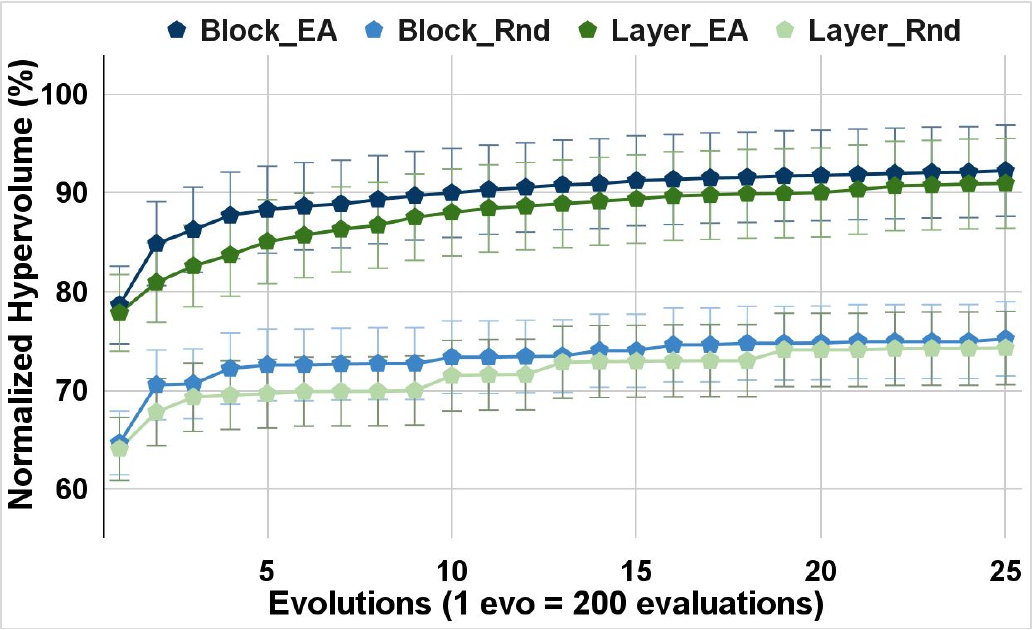}
  \vspace{-3.5mm}
  \caption{Evolutionary Vs. Random Search}
  \vspace{-4mm}
  \label{fig:ea_rnd_maestro}
\end{wrapfigure}

\subsubsection{On the power of evolution}
We further analyze the hypervolume improvement when using an EA compared to a random search. We fix an optimization budget of 5000 evaluations for each and showcase the results in Figure \ref{fig:ea_rnd_maestro} at different evolution stages for the mapping onto 3 CUs experiment. Normalized by a maximum achievable value from our previous results, we observe that the normalized hypervolume in the Figure reaches $\sim$92\% improvement for the EA compared to $\sim$75\% for the random search. We also notice that both \textit{blockwise} and \textit{layerwise} converge to proximate values despite the larger gap at the earlier evolutions (i.e., generations), further indicating the EA's capacity to scale.

\section{Discussion and Future Directions} \label{subsec:future}

\noindent{\large \textcircled{\small 1}} \textbf{Key Takeaways. }Hardware-software design optimizations and workloads mappings are increasingly studied in the literature \cite{fasfous2022anaconga, bouzidi2023hadas, dagli2022axonn, bouzidi2023map}. What distinguishes this work is its specialization in considering the details of: (\emph{i}) GNNs' computational flow irregularity; (\emph{ii}) workload distribution across heterogeneous CUs with varying degrees of support for graph operators. Furthermore, ViG is a relatively emergent class of GNNs, and there remains room for improvement along the design, characterization, and training of ViG supernets, which can only improve as the application of ViGs -- and GNNs in general -- at the Edge continues to proliferate. All things considered, MaGNAS has demonstrated encouraging results that can help pave the way for future lines of research.

\noindent{\large \textcircled{\small 2}} \textbf{Generality and scalability. }In analyzing the generality of MaGNAS (Section \ref{subsec:generality}), we have demonstrated the heterogeneity of hardware accelerators through diversifying dataflows across HW accelerators. In practice, heterogeneity can also occur through varying other factors such as processing engines per accelerator, shared buffer size, off-chip memory bandwidth, etc., all of which can influence the hardware efficiency of the workloads. MaGNAS has been shown capable of generalizing to the different forms of heterogeneity as it relies on high-level performance characterization that abstracts underlying hardware compositions. Furthermore, experiments on real SoCs with different HW accelerators and levels of heterogeneity from that of the Nvidia Xavier is still needed to corroborate that MaGNAS can scale effectively to diverse platforms. 

\noindent{\large \textcircled{\small 3}} \textbf{Graph operation support limitations. }As MAESTRO does not natively support the sparse matrix multiplications, we implemented GNN operations within the simulator as generic matrix multiplications, which has led to considerable execution overheads for the \textit{aggregation} phase regarding latency and energy. This is indeed a situation akin to the case when GNN workloads are to be run on generic, uncustomized edge devices that lack proper support for specialized accelerators for GNN operations. In such cases, mapping optimizations can be particularly beneficial in mitigating the impact of such hardware deficiencies. Furthermore, as GNNs grow in popularity, promising steps are being taken towards developing new dataflows for reconfigurable spatial accelerators to support irregular graph computational sequences, which will also bring about the need for new architectural simulators to effectively model their performance overheads.

\noindent{\large \textcircled{\small 4}} \textbf{Other Application Domains. }Vision-based applications provided practical, tangible use case motivations for the \textit{GNNs-on-SoCs} scenario, and accordingly, they have become the primary target application of this work. With that being said, the manner in which MaGNAS has been developed enables it to generalize to other emerging applications on edge SoCs that employ GNNs for their primary computational workloads. For instance, the support for mapping on both the blockwise and layerwise levels of granularity within MaGNAS enables it, with some fine-tuning, to serve other types of emerging GNN-based applications at the edge by maximizing GNNs' efficiency across a broad range of diverse CUs integrated onto the same chip.

\section{Related Works}\label{sec:related_works}

\noindent{\large \textcircled{\small 1}} \textbf{GNNs for vision.} 
Through learning graph-level features, GNNs achieved remarkable performance on a variety of computer vision tasks, such as activity recognition \cite{yan2018spatial} and point clouds classification \cite{landrieu2018large, wang2019dynamic}. Traditionally, the success of GCNs in computer vision applications relied on the graph construction technique, which in many cases was tailored to suit the input data semantics and downstream task. Scene graph generation \cite{malawade2022spatiotemporal, xu2017scene, yu2021scene} emerged as a viable approach to generate a graph of objects and their relations from an image through cascading an object detector and a GCN model. The ViG \cite{han2022vision}, a generic architecture upon which our framework is constructed, represents a standard GCN backbone to generate and process graphs from raw images to serve general computer vision applications.

\noindent{\large \textcircled{\small 2}} \textbf{Hardware acceleration for GNNs}
The two phases of GNN favor different classes of accelerators: GNN acceleration favors MIMD architectures to address the irregularity of graph operations by providing high random access memory bandwidth and small data access sizes, whereas DNN acceleration is achieved through SIMD architectures for exploiting data locality through caches or local scratchpads. As such, numerous works \cite{yan2020hygcn, auten2020hardware, you2022gcod, stevens2021gnnerator, chen2021dygnn, kiningham2022grip} have proposed hybrid accelerator architectures comprising separate engines and specialized hardware components to effectively manage the non-uniform GNN dataflow on both an \textit{intra-} and \textit{inter-phase} level. However, such proposed accelerator designs are acutely specialized ASICs, complicating their integration into numerous commodity hardware platforms and SoCs. Since GNNs are becoming increasingly popular, recent research efforts \cite{garg2022understanding} have directed their approach towards characterizing the design space of dataflow choices to enable running GNNs on customary reconfigurable spatial accelerators, intending to identify convenient dataflows to service various GNN use cases. The philosophy behind our method follows the latter trend. However, it is complementary to both approaches since it abstracts the underlying accelerator architecture and adds another layer of design space exploration to characterize joint search space of GNN architectures and the inter-phase pipelining across heterogeneous computing components in an SoC.

\noindent{\large \textcircled{\small 3}} \textbf{Distributed Computing of GNNs.} Distributing DNN workloads across the heterogeneous computing resources of CPU, GPU, DLAs, and FPGAs, is an active field of research \cite{dagli2022axonn, bouzidi2023map, pujol2019generating, kim2022energy, xun2020optimising}. Researchers have recently explored how to distribute GNN workloads to enhance performance efficiency by exploiting the underlying heterogeneous hardware composition via task-level, data-level, and pipelining forms of parallelism \cite{chen2022survey}. For instance, the work in \cite{zhang2022low} proposed to decouple GNNs onto CPU-FPGA heterogeneous platform to speedup GNN inference.

\begin{table}[h]
\centering
\caption{Comparison between related Graph Neural Architecture Search works and ours.}
\vspace{-2ex}
\fontsize{9}{9}\selectfont
\scalebox{1.}{
\label{tab:comparison}
\begin{tabular}{l|c c c c c c c}
        \hline
         & \cite{gao2019graphnas} & \cite{gao2020graph} & \cite{zhou2019auto} & \cite{shi2022genetic} & \cite{zhang2021g} & \cite{zhou2023hardware} & MaGNAS \\
        \hline
        Training-in-the-loop NAS & \checkmark & \checkmark & \checkmark & \checkmark & & & \\
        Once-for-all NAS & & & & & \checkmark & \checkmark & \checkmark\\
        Vision GNN & & & & & & & \checkmark \\
        Hardware Awareness & & & & & \checkmark & \checkmark & \checkmark \\
        GNN-Hardware co-design & & & & & \checkmark & & \\
        Edge Computing Setting & & & & & & \checkmark & \checkmark \\
        Distributed Mapping & & & & & & & \checkmark \\
        \hline
        \hline       
    \end{tabular}
}
\end{table}

\noindent{\large \textcircled{\small 4}} \textbf{Graph Neural Architecture Search.} Recent research works investigated how to leverage the power of Neural Architecture Search to automate the design process of GNNs. Earlier works adopted search approaches like Reinforcement Learning \cite{gao2019graphnas, gao2020graph, zhou2019auto} or Evolutionary algorithms \cite{shi2022genetic}. The work in \cite{you2020design} further proposed a generalized GNNs' design space with a knowledge distillation method from GNN model-task pairs. 
However, these approaches mostly fall under the training-in-the-loop NAS category.
Furthermore, limited or no awareness of the underlying hardware computing platform capabilities was taken. As such, more recent works in \cite{zhang2021g, zhou2023hardware} proposed to move towards the once-for-all approach \cite{cai2019once}, which 
employs a supernet that characterizes the design space of the GNN architectures. Specifically, the training of the supernet can be conducted only once by leveraging the property of weight-sharing. 
On the hardware side, \cite{zhang2021g} adopts a co-design NAS approach for GNN and hardware accelerator, whereas \cite{zhou2023hardware} optimizes the GNN design to suit underlying commodity edge computing platforms. Our work falls under the same category of HW-aware NAS for GNNs as these two. However, several features distinguish this work from others: (\emph{i}) our supernet is designed to consider the emerging class of vision-based GNNs (ViGs); 
(\emph{ii}) support for evaluating candidate ViG subnets during the search process based on their best mapping options that leverage pipelining parallelism across diverse computing units within the MPSoC edge platform; 
(\emph{iii}) our two-tier search algorithm implementation allows the inner optimization engine to be extensible to other MPSoCs and GNN supernets serving other tasks. We summarize the differences in Table \ref{tab:comparison}.

\section{Conclusion}

In this paper, we presented MaGNAS, a mapping-aware Graph Neural Architecture Search framework for the distributed deployment of vision GNN onto heterogeneous SoCs. MaGNAS characterizes a GNN architectural design space bound with prospective mapping options, enabling the identification of model designs optimized to the distributed deployment scheme. MaGNAS employs a two-tier evolutionary search framework to identify optimal \textit{architecture} and \textit{mapping} pairings that provide the best performance trade-offs. Extensive experimentation, in-depth analysis, and ablation studies using a real MPSoC platform and hardware simulation have showcased the merit of MaGNAS in designing ViG architectures and mapping them onto heterogeneous MPSoCs.

\section*{Acknowledgement}
This work was supported by the National Science Foundation (NSF) under award CCF-2140154.

\bibliographystyle{ACM-Reference-Format}
\bibliography{references}

\end{document}